\documentclass[final,10pt,times,onecolumn]{elsarticle}
\usepackage[margin=1.8cm]{geometry}
\usepackage{amsmath}
\usepackage{amsfonts}
\usepackage{amssymb}
\usepackage{graphicx}
\usepackage{lmodern}
\usepackage{hyperref}
\usepackage{dcolumn}
\usepackage{bm}
\usepackage[mathlines]{lineno}
\usepackage{float}
\usepackage[T1]{fontenc}
\usepackage{subfigure}
\usepackage{mathtools, cuted}
\usepackage[english]{babel}
\usepackage{lipsum}

\journal{Physics Letters B}

\usepackage{soul}
\usepackage{xcolor}

\usepackage{numcompress}

\DeclareUnicodeCharacter{2212}{\ensuremath{-}}

\begin{document}
\newcolumntype{P}[1]{>{\centering\arraybackslash}p{#1}}
\title{\textbf{Neutrino Mixing from a Fresh Perspective}}



\author[mymainaddress]{Pralay Chakraborty}
\ead{pralay@gauhati.ac.in}

\author[mymainaddress]{Manash Dey}
\ead{manashdey@gauhati.ac.in}

\author[hismainaddress]{Biswajit Karmakar}
\ead{biswajit.karmakar@us.edu.pl}

\author[mymainaddress]{Subhankar Roy\corref{mycorrespondingauthor}}
\cortext[mycorrespondingauthor]{Corresponding author}
\ead{subhankar@gauhati.ac.in}

\address[mymainaddress]{Department of Physics, Gauhati University, India}

\address[hismainaddress]{Institute of Physics, University of Silesia, Katowice, Poland}

\begin{abstract}
We propose a neutrino mass matrix texture bearing a suitable correlation $m_{22}=-2\,m_{13}$ and study its phenomenological implications. In light of both normal and inverted hierarchies, the texture imposes specific bounds on some observational parameters. As a potential application, the prediction of effective Majorana neutrino mass $m_{\beta\beta}$ is visualized for both hierarchies. To understand the proposed texture from first principle, we incorporate type-I+II seesaw mechanism in association with $A_4 \times Z_{10} \times Z_2$ group.
\end{abstract}
\maketitle

\section{Introduction \label{section 1}}

In the realm of particle physics, neutrinos are mysterious creatures that can transform from one type to another as they traverse space. This phenomenon is called neutrino oscillation and this idea is first introduced by Bruno Pontecorvo in the year 1957 \cite{Pontecorvo:1957cp}. The neutrino oscillation arises from the fact that the three neutrino flavour states ($\nu_{l=e,\mu,\tau}$) are admixtures of three mass eigenstates $(\nu_{i=1,2,3})$ with corresponding mass values ($m_{i=1,2,3}$). The Pontecorvo-Maki-Nakagawa-Sakata (PMNS) matrix\,\cite{Maki:1962mu} is the mathematical representation that connects the flavour states to the mass states, providing a framework for understanding how neutrinos transform between different flavours as they propagate through space. The neutrino oscillation demands the existence of a non-zero neutrino mass, contradicting the predictions of the Standard Model\,(SM)\,\cite{Glashow:1961tr, Weinberg:1967tq, Salam:1968rm}. In this regard, to understand the origin of neutrino mass, it is necessary to explore theories that go beyond the SM. The neutrino mass matrix ($M_\nu$) originates from the Yukawa Lagrangian, which is formulated within the framework of the seesaw mechanism \cite{Yoshimura:1978ex, Akhmedov:1999tm, Schechter:1980gr, Schechter:1981cv}. In the literature, the widely used seesaw mechanisms are the type-I seesaw\,\cite{Minkowski:1977sc, Gell-Mann:1979vob, Mohapatra:1979ia} and type-II seesaw mechanism\,\cite{Magg:1980ut, Cheng:1980qt, Lazarides:1980nt, Mohapatra:1980yp}. 

The Majorana neutrino mass matrix $M_\nu$ contains twelve parameters: three mass eigenvalues ($m_{i=1,2,3}$), three mixing angles ($\theta_{12}, \theta_{13}, \theta_{23}$), three Charge Parity\,(CP) violating phases ($\delta, \alpha, \beta$), and three unphysical phases ($\phi_1, \phi_2, \phi_3$). In recent years, the experiments predict the mixing angles with high precision but the precise determination of Dirac CP phase $\delta$ and the octant of the atmospheric mixing angle $\theta_{23}$ yet to be settled \cite{Gonzalez-Garcia:2021dve}. In addition, the experiments are unable to provide exact values of individual mass eigenvalues; instead, they offer predictions for two mass-squared differences. In addition, the Majorana phases are physical parameters that are still untouched in the experiments. However, as discussed above, the neutrino mass matrix carries information on twelve parameters. By imposing certain constraints on the neutrino mass matrix, some of the physical parameters can be predicted. In the literature, various such constraints are studied: texture zeroes \cite{Ludl:2014axa}, vanishing minors \cite{Lashin:2009yd}, hybrid textures \cite{Liu:2013oxa}, and $\mu-\tau$ symmetry \cite{Harrison:2002er}. Due to an inherent connection with the $\mu-\tau$ symmetry, several mixing patterns are studied in the literature. Among them, bimaximal mixing\,(BM)\,\cite{Barger:1998ta}, tri-bimaximal mixing\,(TBM)\,\cite{Harrison:2002er}, hexagonal mixing\,(HM)\,\cite{Albright:2010ap} and golden ratio\,(GR)\,\cite{Ding:2011cm} are notable. The study of these mixing schemes holds significance from the model-building perspective \,\cite{Xing:2020ijf, Ma:2005qf}. In this regard, the tri-bimaximal mixing has been investigated within the framework of various discrete flavour symmetries\,\cite{Ma:2004zv, Altarelli:2005yp}, for a recent review on discrete flavour symmetries and their implication in neutrino mixing, see Ref.\,\cite{Chauhan:2023faf} and for specific examples, see Refs.~\cite{Karmakar:2014dva, Borah:2017dmk} and references there in. However, present experiments have ruled out $\mu-\tau$ symmetry as it predicts a vanishing $\theta_{13}$. Apart from the traditional mixing schemes discussed above, our group has proposed a new fixed mixing scheme, as shown in Ref.\,\cite{Dey:2023rht}. 

In principle, one can start from $\mu$-$\tau$ symmetry and explore the deviations from it through different mechanisms. Alternatively, the neutrino mass matrix can be treated from a fresh perspective without considering the signs of $\mu$-$\tau$ symmetry \cite{Dey:2022qpu, Chakraborty:2023msb, Chakraborty:2024rgt}. A neutrino mass matrix contains all the information of observational parameters. If we demand the neutrino mass matrix to be predictive, then the matrix elements must, somehow, become correlated. For example, in the "Texture Zero" approach, simple correlations are drawn by setting some of the matrix elements to zero. However, in our approach, we explore the phenomenological consequences of imposing certain constraints on the neutrino mass matrix elements \cite{Chakraborty:2022ess, Chakraborty:2024hhq}. With this motivation, we propose a simple correlation as shown below:

\begin{eqnarray}
\label{correlation}
  m_{22}&=&-2\,m_{13},
\end{eqnarray} 
where $m_{ij}$'s are the neutrino mass matrix elements. Assuming the neutrinos as Majorana fermions, the proposed neutrino mass matrix appears as,  

\begin{equation}
M_\nu=\begin{bmatrix}
A & B & F\\
B & -2F & G\\
F & G & J
\end{bmatrix},
\label{eq:mass-mat-textute}
\end{equation}
where, $A,B,F,G$ and $J$ are, in general, complex parameters. Needless to mention, the proposed texture is expected to align with the known experimental data. The above texture exhibits certain interesting features. For example, it restricts the predictions of the atmospheric mixing angle. Similarly, the three CP-violating phases are also constrained within certain bounds. Though it cannot discriminate between normal and inverted hierarchies of neutrino masses, it supports a quasi-degenerate spectrum of neutrino masses for the latter scenario. However, the proposed texture cannot constrain the reactor and solar angles. In the present work, we have studied the proposed correlation in Eq.\,(\ref{correlation}) in the context of Majorana fermions. The consequences of the Dirac case remain unexplored and present a prospect for future research. It may appear that the mass matrix shown in Eq.,(\ref{eq:mass-mat-textute}) is arbitrary. Now, it is important to find if discrete flavor symmetric frameworks, often adopted to explain the fermionic flavor structure,  can reproduce such texture, providing a novel theoretical origin. Interestingly, here we find that the proposed texture can be realized in its exact form starting from a seesaw model based on the $A_4 \times Z_{10} \times Z_2$ group.

The plan of the paper is outlined as follows: In Section (\ref{section 2}), we give the formalism of the work. The section (\ref{section 3}) is devoted towards numerical analysis and predictability of the proposed texture. The predictions involve neutrino mass eigenvalues, associated Majorana phases, and effective mass parameters appearing in the neutrinoless double beta decay. The section (\ref{section 4}) is devoted to the model associated to the proposed texture. Finally, we summarize and conclude in Section (\ref{section 5}).

\section{Formalism \label{section 2}}

The Majorana neutrino mass matrix being complex symmetric in nature, requires a unitary matrix to get diagonalised. In the present work, we use the parametrization of the unitary matrix $U$ as provided by the Particle Data Group (PDG) \cite{ParticleDataGroup:2020ssz},

\begin{eqnarray}
U &=& \begin{bmatrix}
e^{i\phi_1} & 0 & 0\\
0 & e^{i\phi_2} & 0\\
0 & 0 & e^{i\phi_3}
\end{bmatrix}\times\begin{bmatrix}
1 & 0 & 0\\
0 & c_{23} & s_{23}\\
0 & - s_{23} & c_{23}
\end{bmatrix}\times \begin{bmatrix}
c_{13} & 0 & s_{13}\,e^{-i\delta}\\
0 & 1 & 0\\
-s_{13} e^{i\delta} & 0 & c_{13}
\end{bmatrix} \times \begin{bmatrix}
c_{12} & s_{12} & 0\\
-s_{12} & c_{12} & 0\\
0 & 0 & 1
\end{bmatrix},
\end{eqnarray}

where, $s_{ij}=\sin\theta_{ij}$ and $c_{ij}=\cos\theta_{ij}$. It is to be noted that the unphysical phases $\phi_1$, $\phi_2$ and $\phi_3$ can be absorbed by redefining the charged lepton fields in terms of these phases. In general, the Majorana phases $\alpha$ and $\beta$ are incorporated into the PMNS matrix $U$. However, these phases can also be placed within the diagonal neutrino mass matrix. In the present work, we adopt the latter approach as shown below,

\begin{eqnarray}
M_\nu^{\text{diag}} &=& \begin{bmatrix}
m_1 e^{-2i\alpha} & 0 & 0\\
0 & m_2 e^{-2i\beta} & 0\\
0 & 0 & m_3
\end{bmatrix}.
\end{eqnarray}

We diagonalize the neutrino mass matrix, as shown in Eq.\,(\ref{eq:mass-mat-textute}), using the following transformation,

\begin{equation}
M_\nu^{diag}=U^T M_\nu U.
\end{equation}

 We adopt the said parametrization to ensure that $\alpha$ and $\beta$ can be directly calculated from the following relations,

\begin{eqnarray}
\alpha&=&-\frac{1}{2} Arg\,[M^{diag}_{\nu_{11}}] \label{alpha eq},\\
\beta&=&-\frac{1}{2} Arg\,[M^{diag}_{\nu_{22}}] \label{beta eq}.
\end{eqnarray}

The present analysis involves five complex texture parameters and nine observational parameters. We have three complex diagonalizing conditions appearing as transcendental equations. Additionally, the chosen parametrization ensures that $M_{\nu_{33}}^{diag}$ is real. Therefore, if we exploit the said conditions, we see that some of the free parameters can be reduced. In the next section, we present the numerical analysis for this work.

\section{Numerical Analysis \label{section 3}}

We note that the present analysis includes a large number of free parameters. Instead of fixing the three mixing angles $\theta_{ij}$ and the Dirac CP phase $\delta$ at specific values, we generate a sufficient number of random data points for these parameters, ensuring they are consistent within the $3\sigma$ bound\,\cite{Gonzalez-Garcia:2021dve}. We invoke the conditions as discussed in section\,\ref{section 2} and see that three mass eigenvalues as well as the two Majorana phases can be expressed in terms of Re$[G]$, Re$[J]$ and Im$[J]$. The analysis is performed in such a way that the numerical values of the free parameters Re$[G]$, Re$[J]$ and Im$[J]$ are consistent with the $3\sigma$ bounds of two mass squared differences $\Delta m_{21}^2$ and $\Delta m_{31}^2$\,\cite{Gonzalez-Garcia:2021dve}, while the $\Sigma m_i$ aligns with the Cosmological data\,\cite{Planck:2018vyg}. From Eq.\,(\ref{alpha eq}) and (\ref{beta eq}), we calculate the two Majorana phases $\alpha$ and $\beta$. The analysis is done for both normal and inverted hierarchies.

\begin{figure}
  \centering
    \subfigure[]{\includegraphics[width=0.31\textwidth]{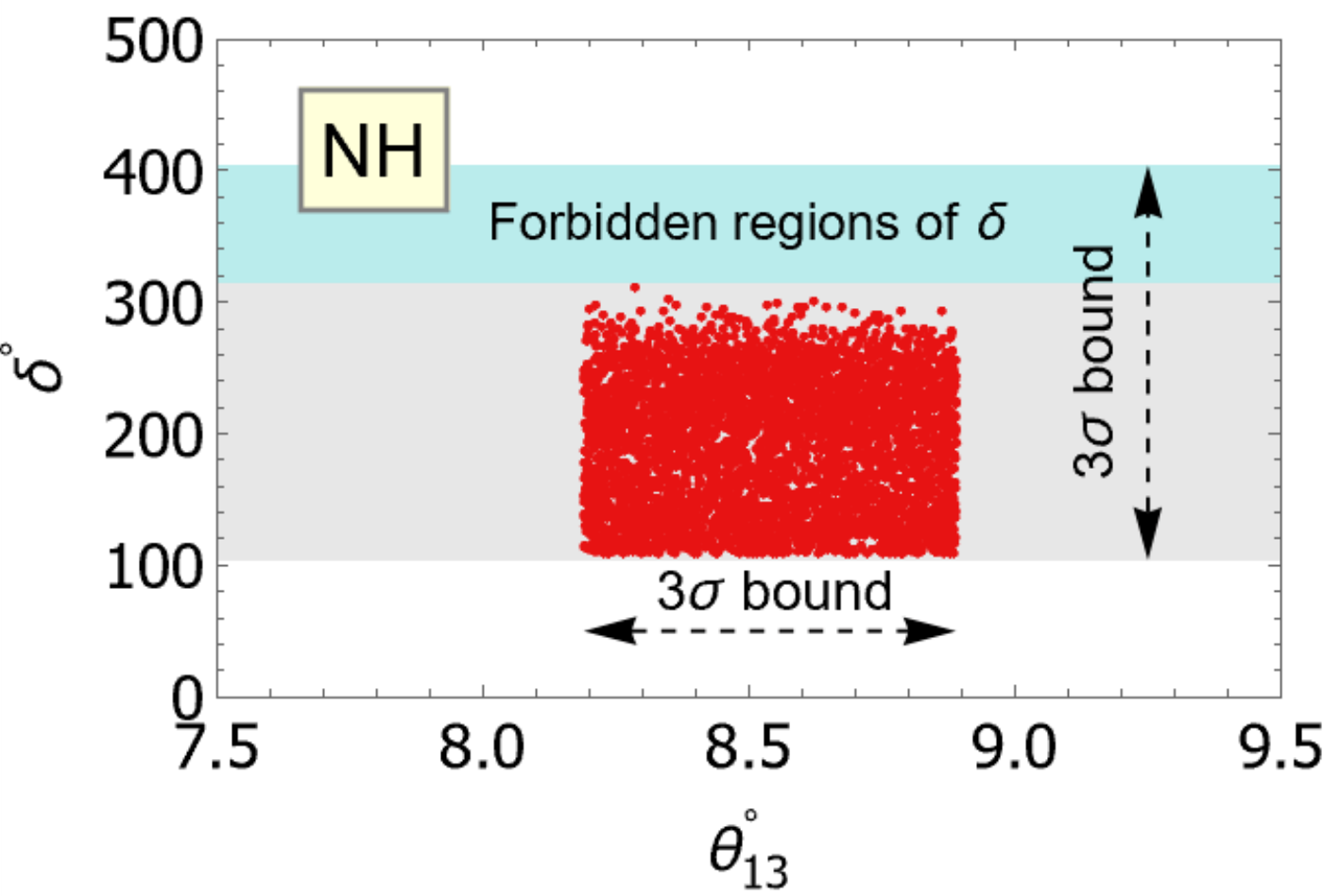}\label{fig:1a}} 
    \subfigure[]{\includegraphics[width=0.31\textwidth]{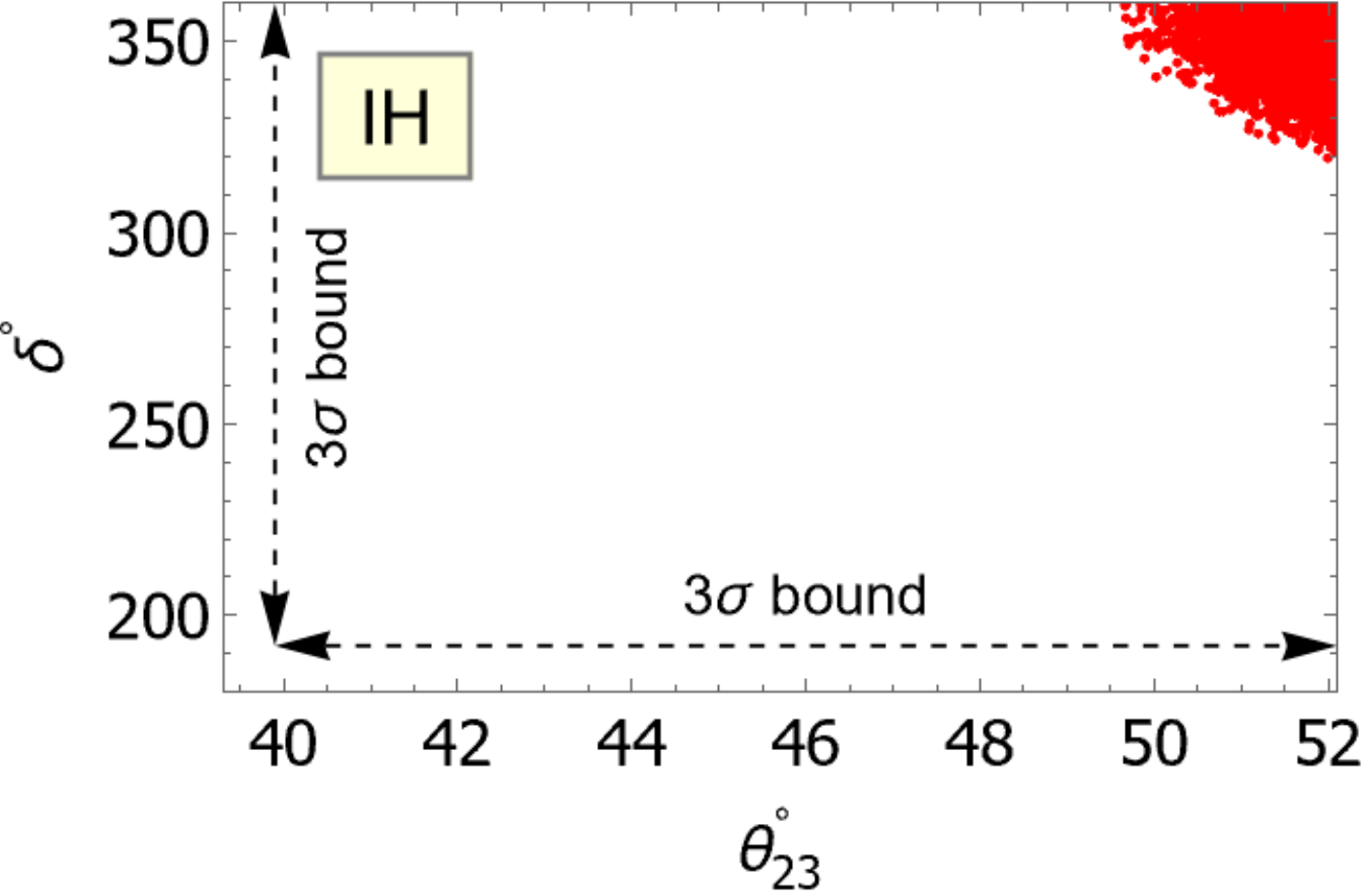}\label{fig:1b}}\\
    \subfigure[]{\includegraphics[width=0.31\textwidth]{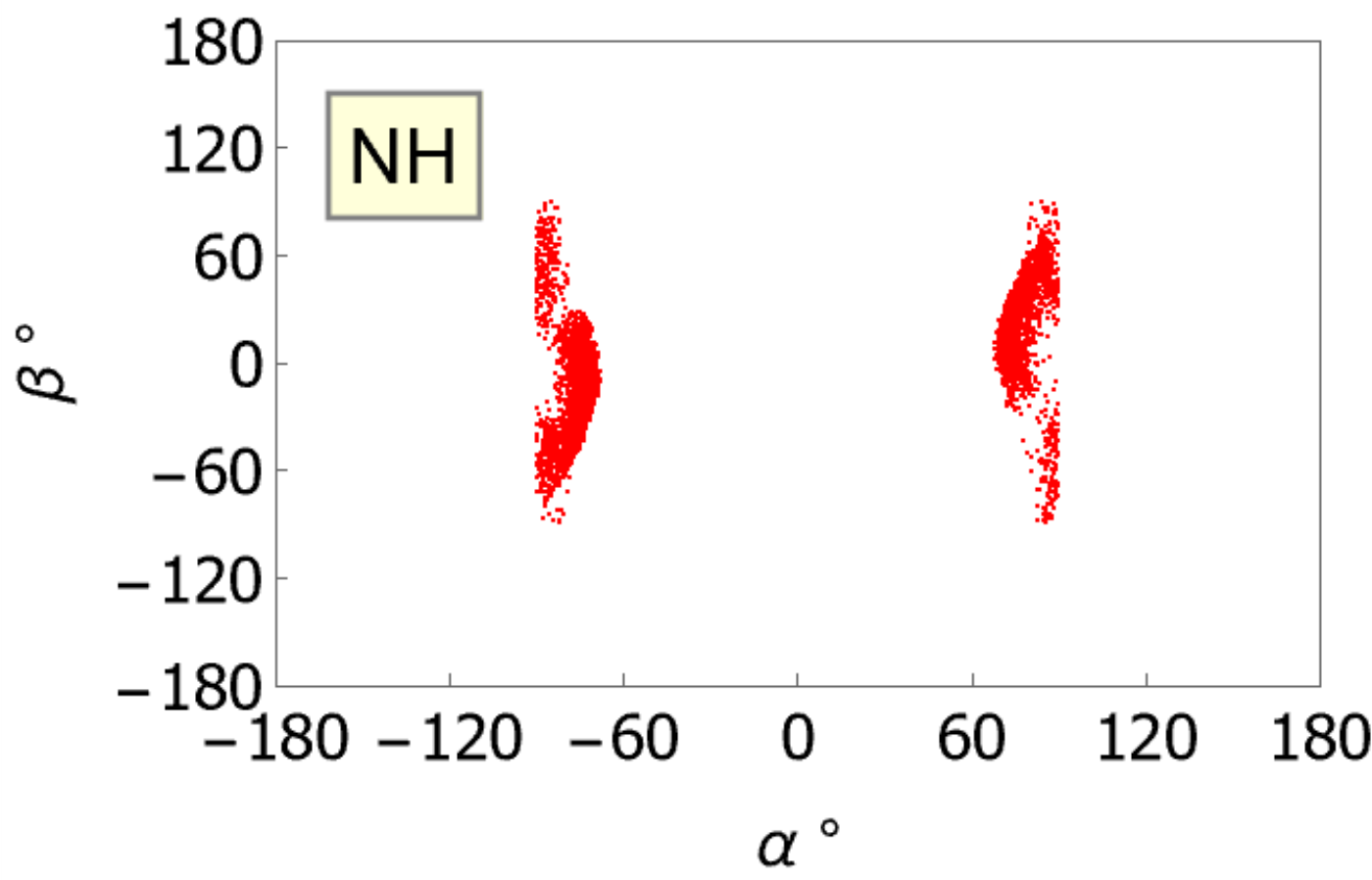}\label{fig:1c}} 
    \subfigure[]{\includegraphics[width=0.31\textwidth]{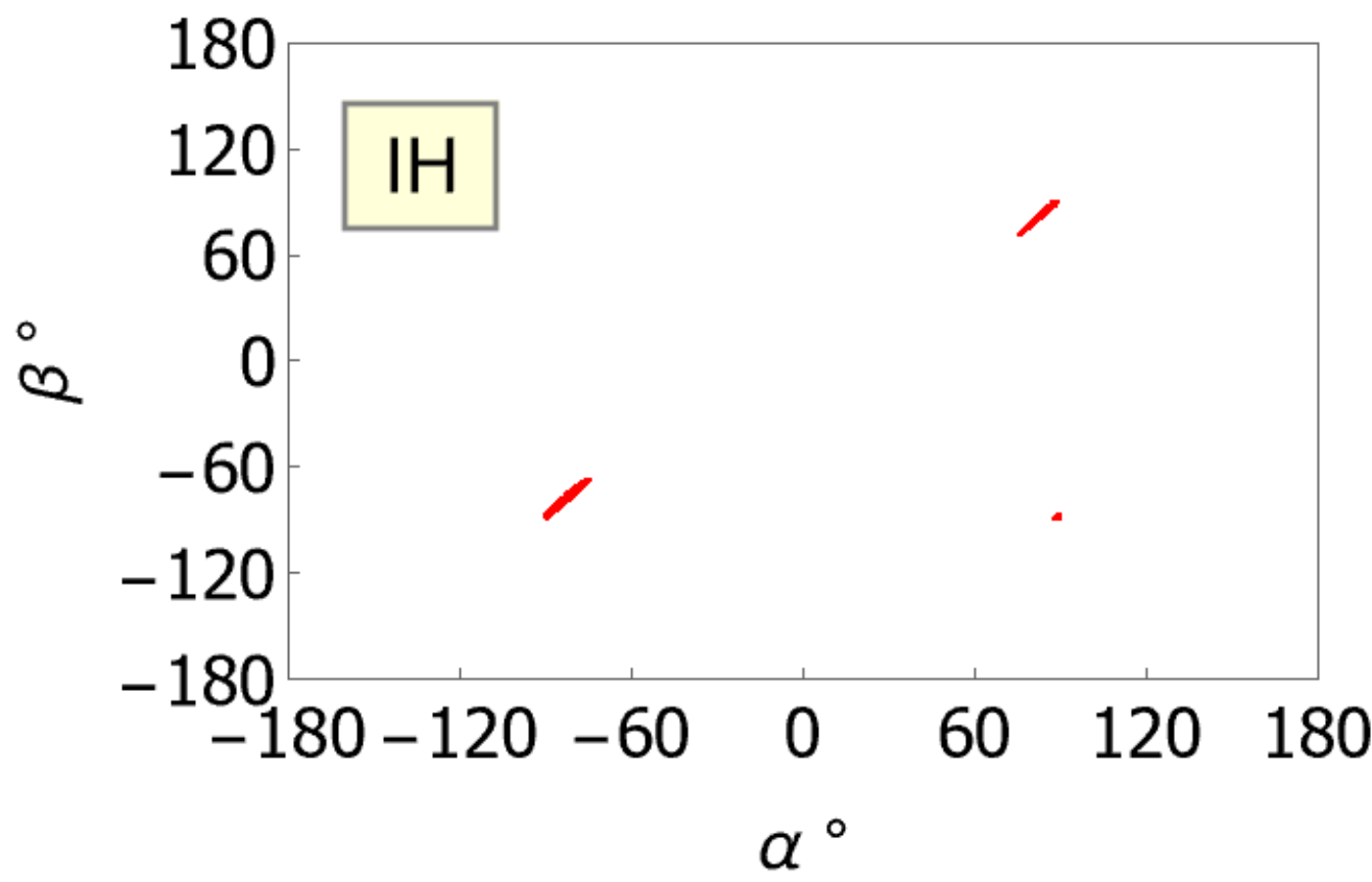}\label{fig:1d}}
   \caption{ (a) The allowed and forbidden values of $\delta$ for normal hierarchy, (b) Shows the allowed and forbidden values of $\theta_{23}$ and $\delta$ for inverted hierarchy, (C) Shows the correlation plots of $\alpha$ vs $\beta$ for the normal hierarchy, (d) Shows the correlation plots of $\alpha$ vs $\beta$ for the inverted hierarchy.}
\label{fig:physical parameters}
\end{figure}

\begin{figure}
  \centering
    \subfigure[]{\includegraphics[width=0.3\textwidth]{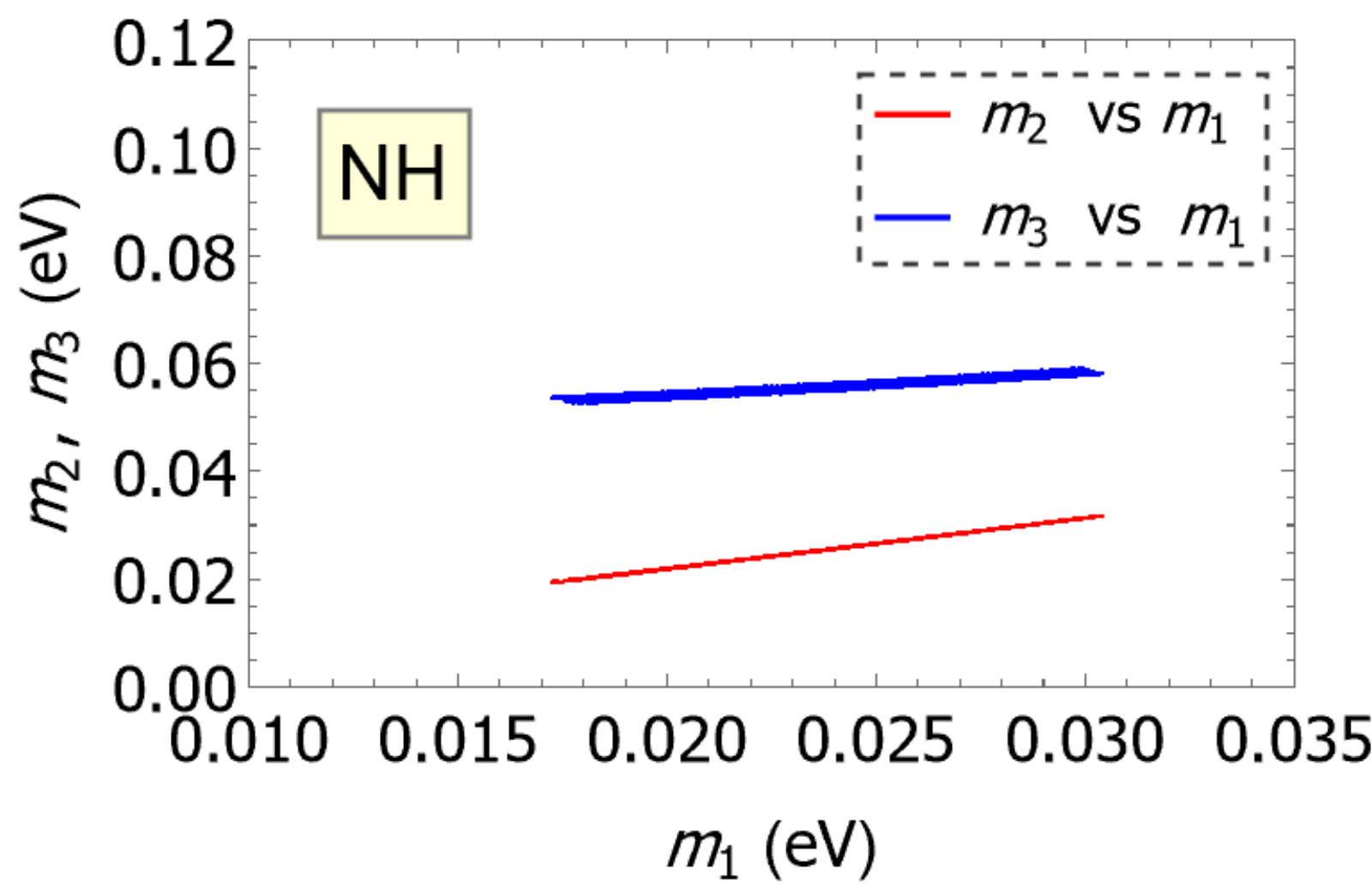}\label{fig:2a}} 
    \subfigure[]{\includegraphics[width=0.3\textwidth]{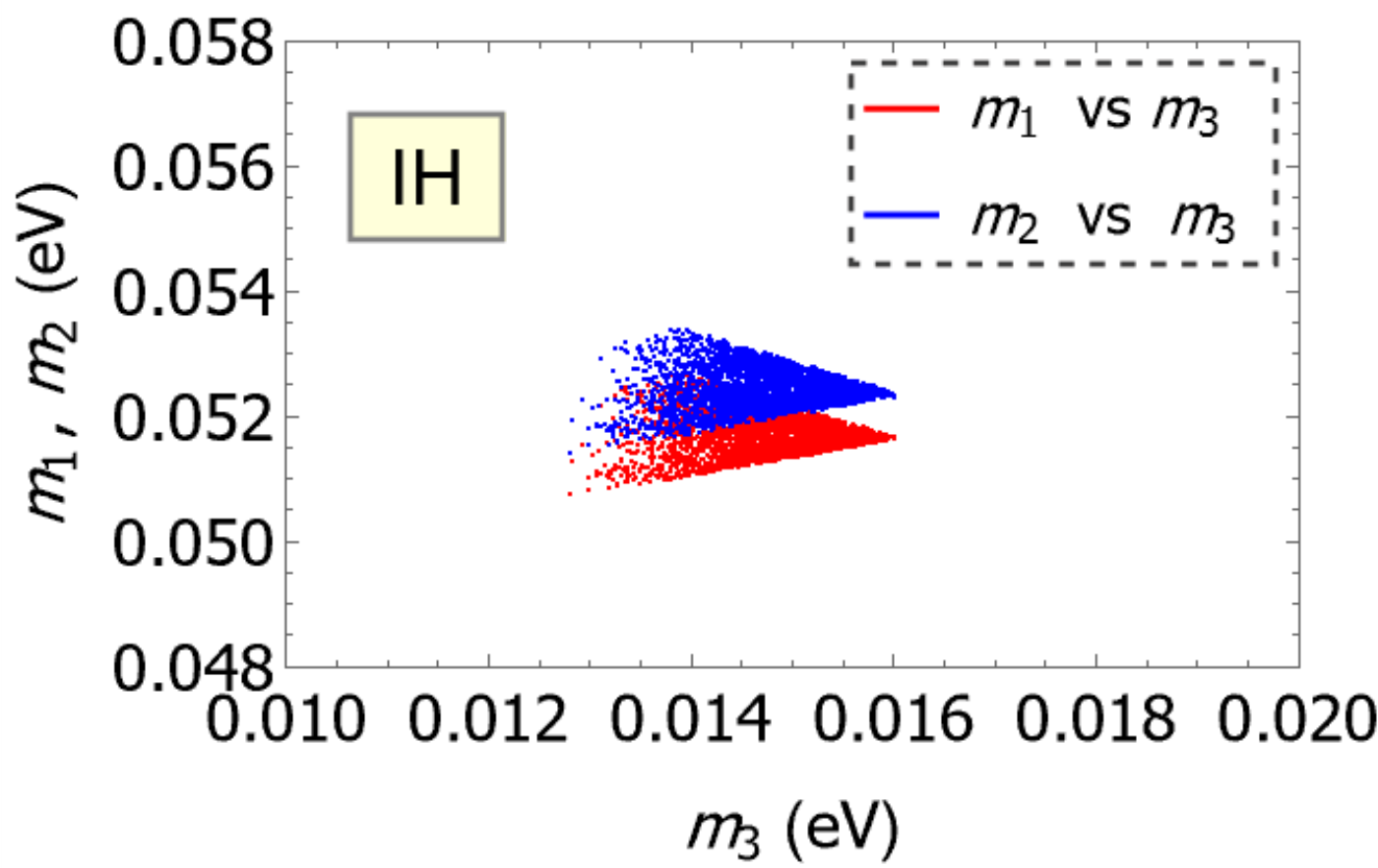}\label{fig:2b}}
    \subfigure[]{\includegraphics[width=0.3\textwidth]{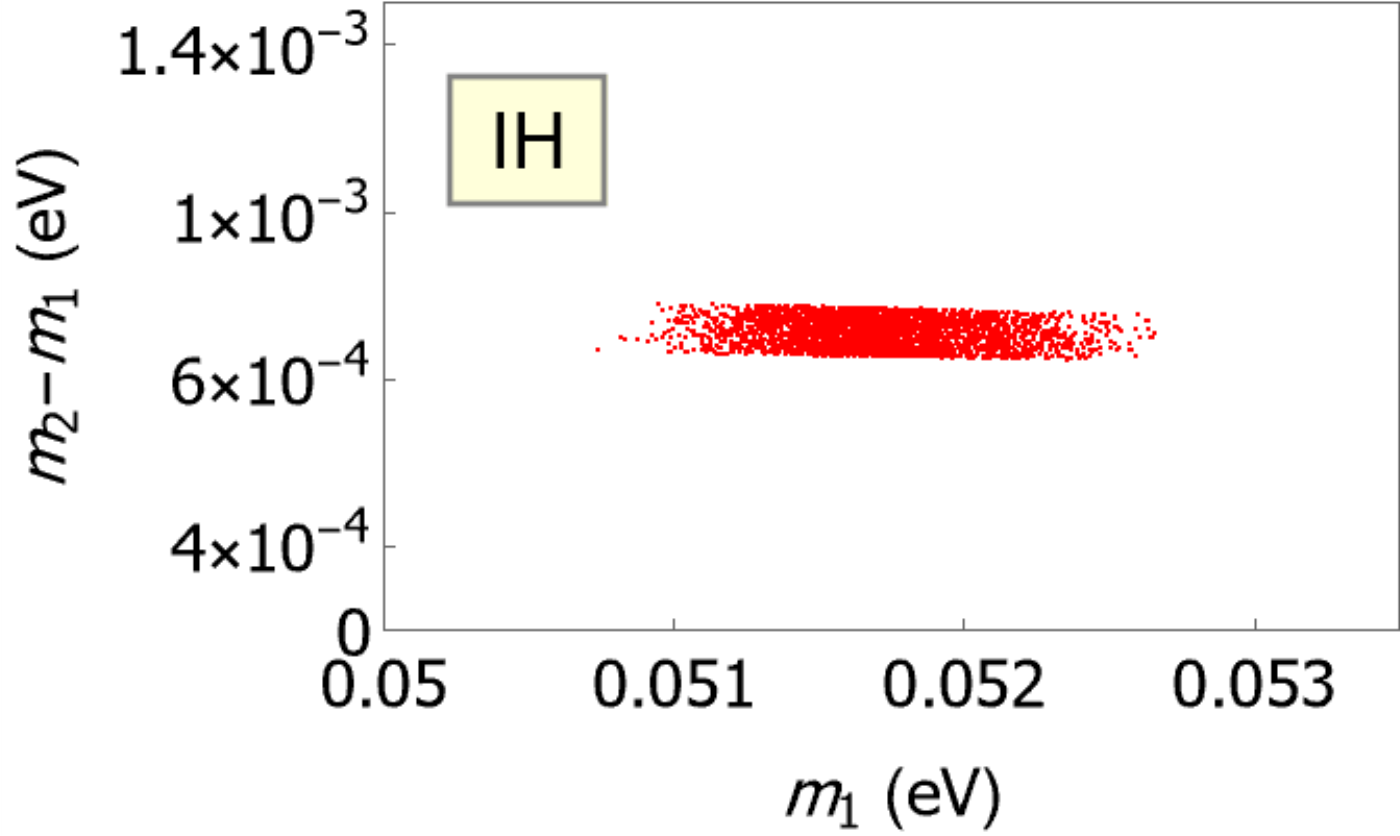}\label{fig:2c}} 
   \caption{ The correlation plots between (a) $m_1$ vs $m_2$ and $m_3$ for normal hierarchy, (b) $m_3$ vs $m_2$ and $m_1$ for inverted hierarchy, (c) $m_2 - m_1$ vs $m_1$ for inverted hierarchy. }
\label{fig:mass parameters}
\end{figure}

\begin{figure}
  \centering
    \subfigure[]{\includegraphics[width=0.33\textwidth]{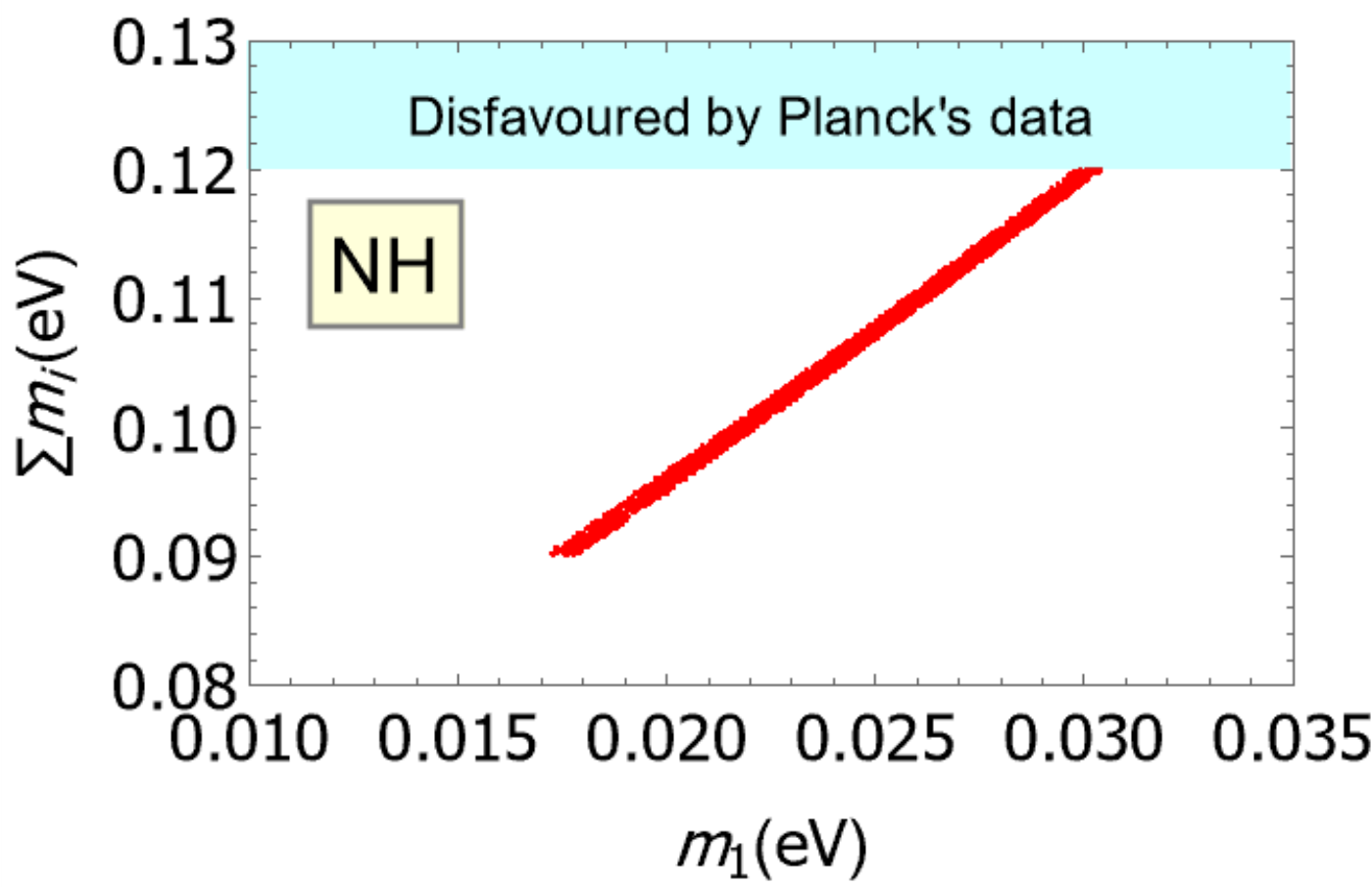}\label{fig:3a}} 
    \subfigure[]{\includegraphics[width=0.33\textwidth]{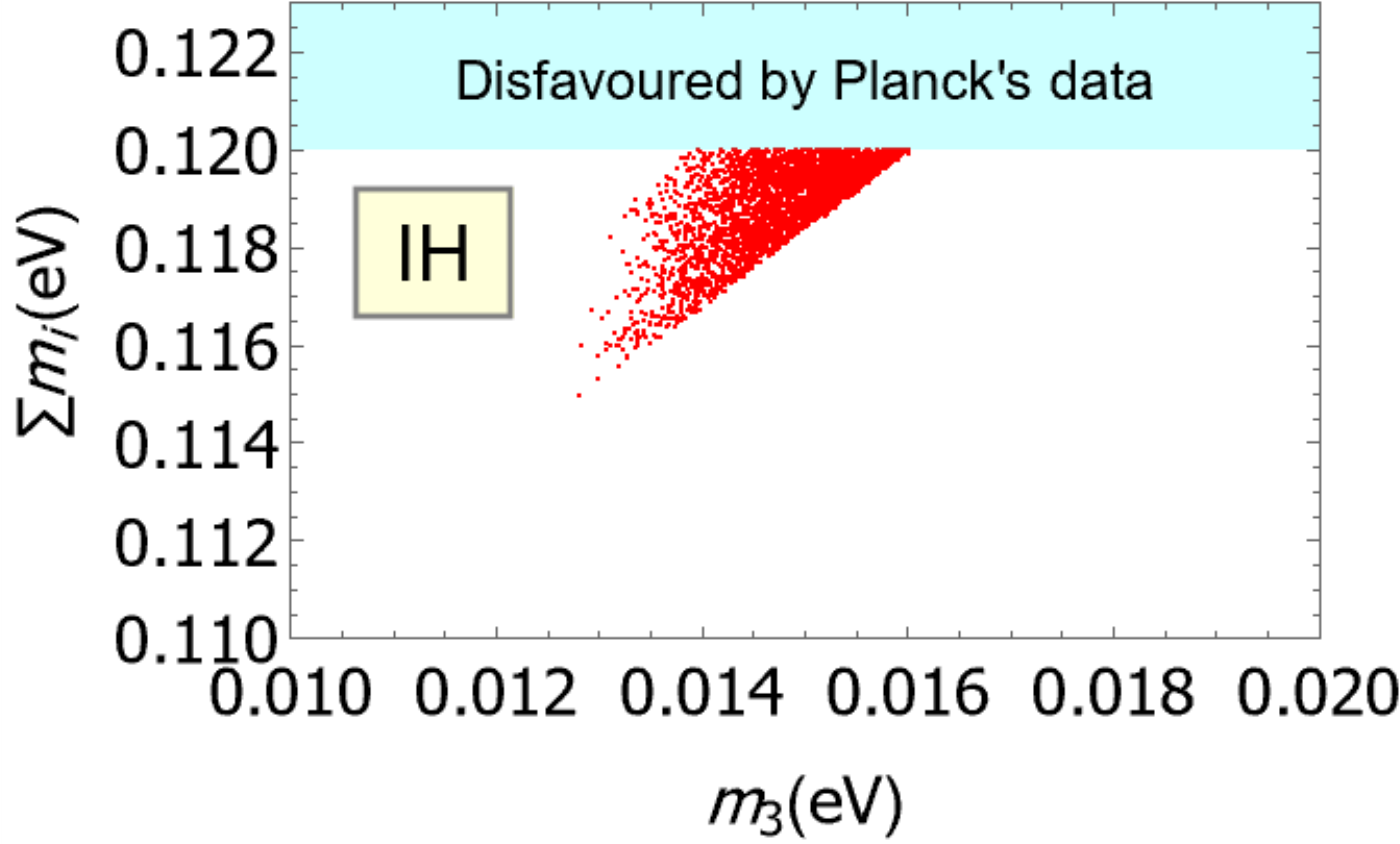}\label{fig:3b}}  
   \caption{ The correlation plots between (a) $m_1$ vs $\Sigma m_i$ for normal hierarchy, (b) $m_3$ vs $\Sigma m_i$ for inverted hierarchy.}
\label{fig:sum}
\end{figure}

\begin{figure}
  \centering
    \subfigure[]{\includegraphics[width=0.33\textwidth]{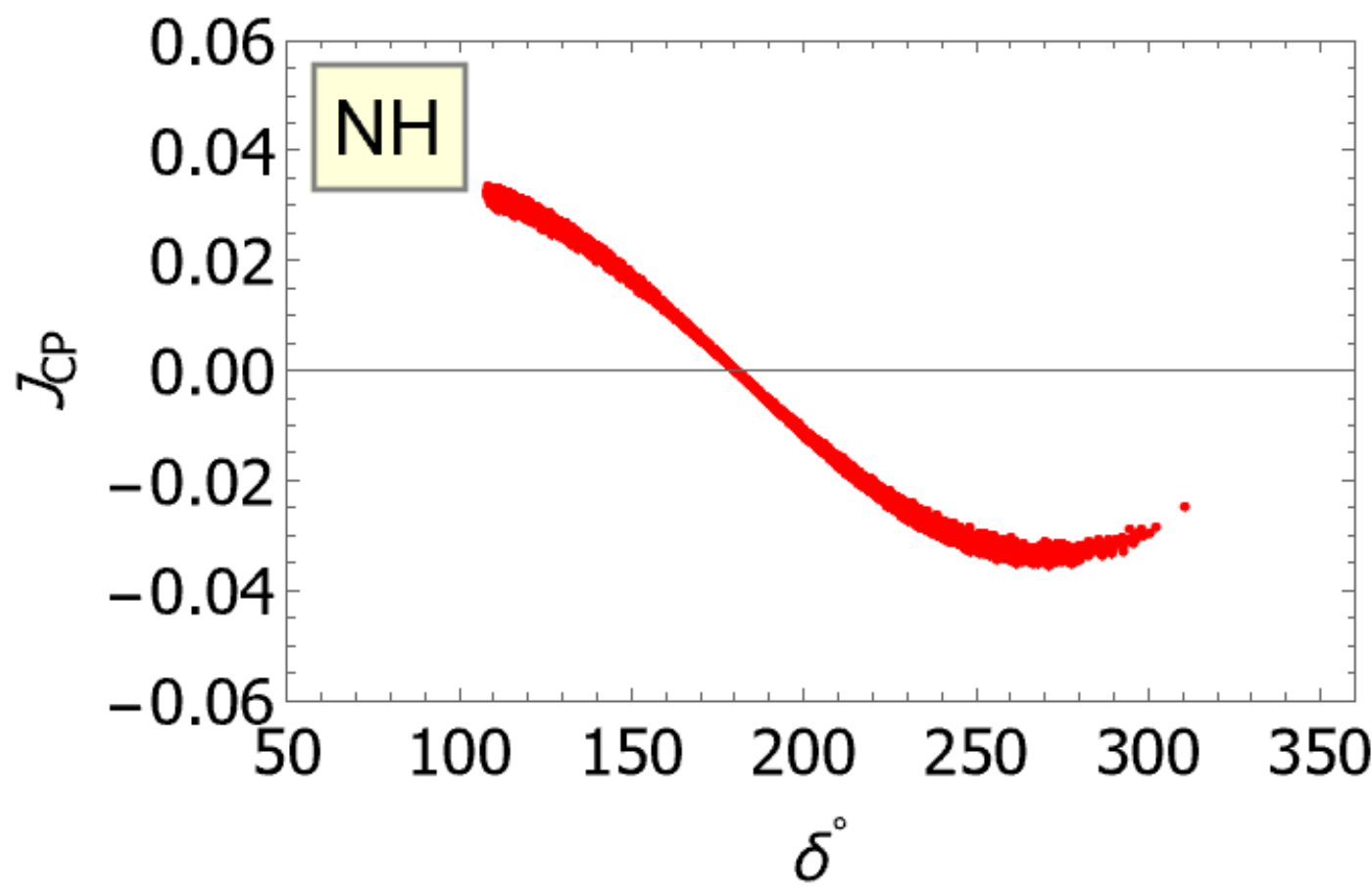}\label{fig:4a}} 
    \subfigure[]{\includegraphics[width=0.33\textwidth]{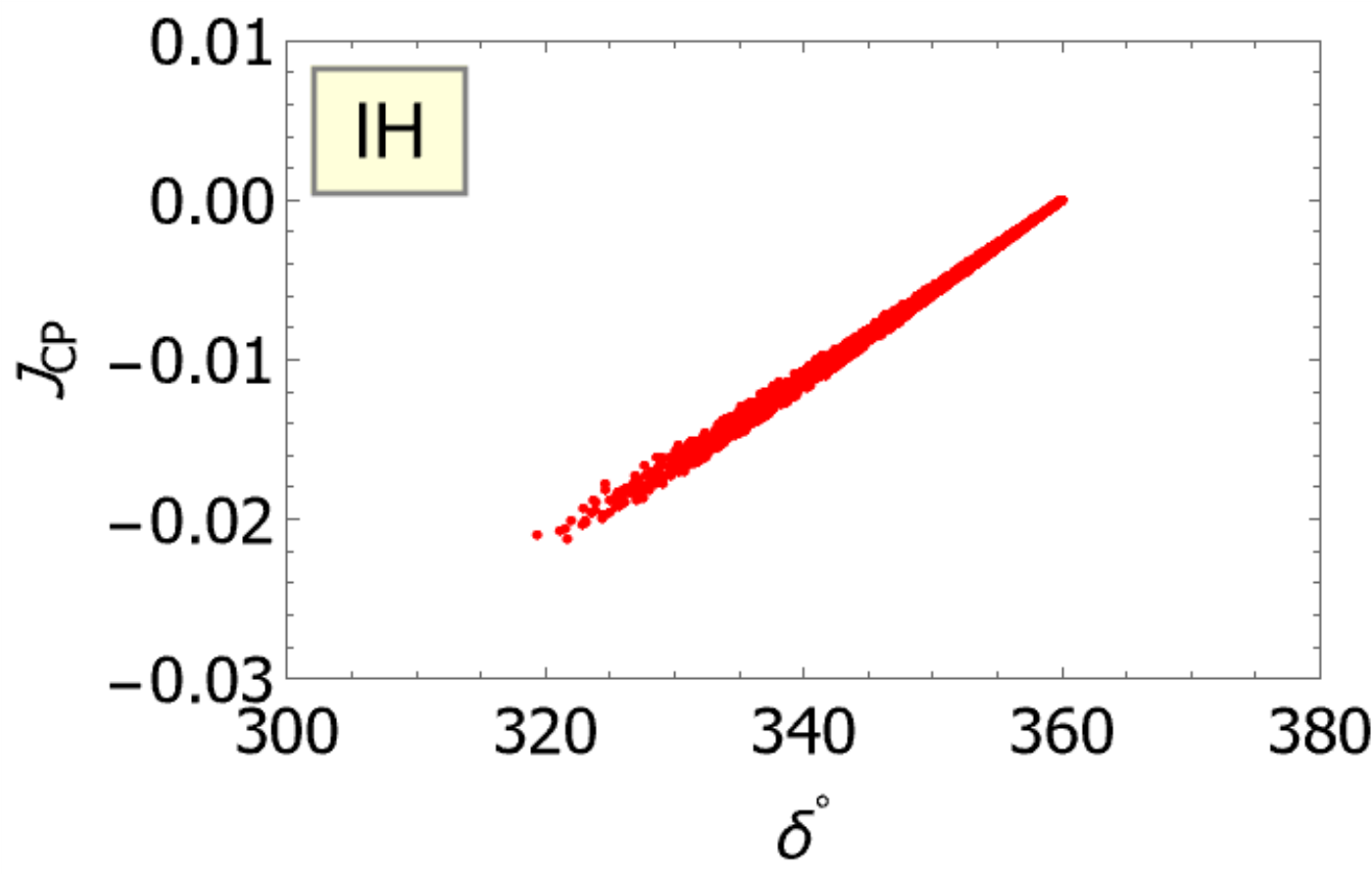}\label{fig:4b}}  
   \caption{ The correlation plots between (a) $J_{CP}$ vs $\delta$ for normal hierarchy, (b) $J_{CP}$ vs $\delta$ for inverted hierarchy.}
\label{fig: jcp vs delta}
\end{figure}

\begin{figure}
  \centering
    \subfigure[]{\includegraphics[width=0.31\textwidth]{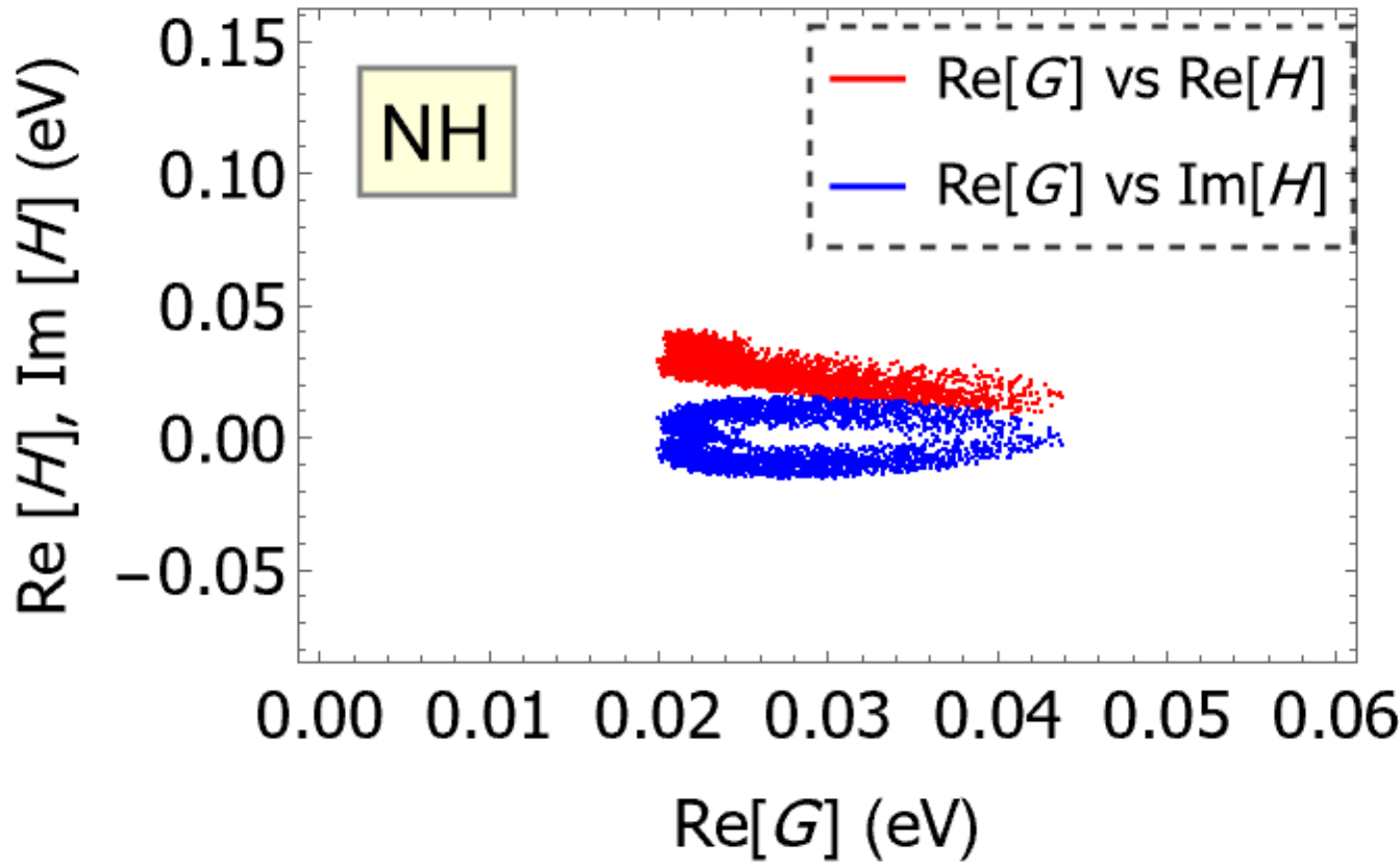}\label{fig:5a}} 
    \subfigure[]{\includegraphics[width=0.31\textwidth]{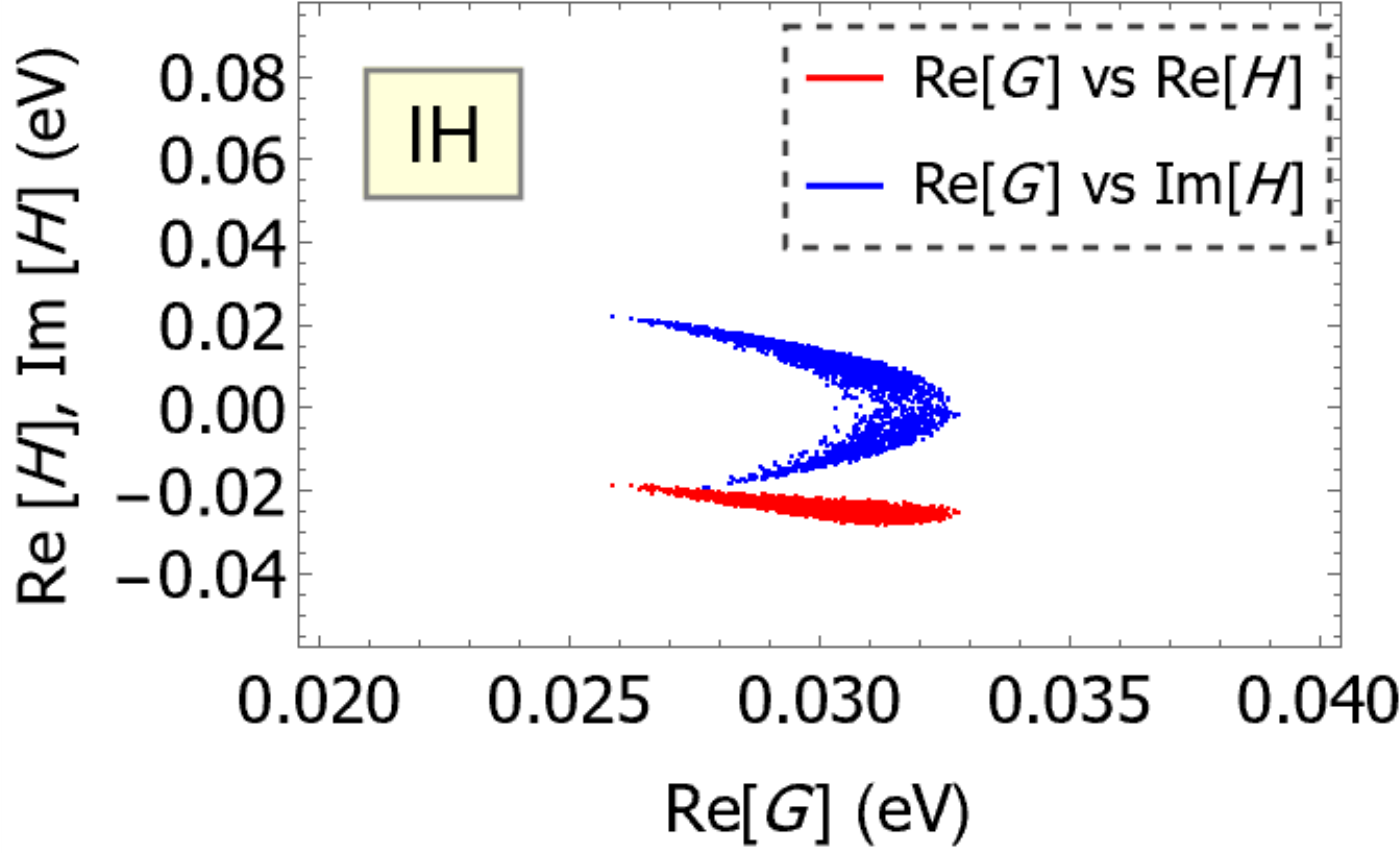}\label{fig:5b}}
   \caption{ The correlation plots between Re[$J$] and Im[$J$] vs Re[$G$] for (a) Normal hierarchy, (b) Inverted hierarchy.}
\label{fig:texture parameters}
\end{figure}

\newcolumntype{P}[1]{>{\centering\arraybackslash}p{#1}}
\begin{table*}
\centering
\begin{tabular} {c c c}
\hline
Parameters & Minimum Value & Maximum Value\\
\hline
\hline
$m_1 /\text{eV}$ & 0.0172 & 0.0304\\
\hline
$m_2 /\text{eV}$ & 0.01922 & 0.0316 \\
\hline
$m_3 /\text{eV}$ & 0.0524 & 0.059\\
\hline
$\sum{m_i} /\text{eV}$ &0.09 &0.119\\
\hline
$|\alpha|/^\circ$ & $68.40$ & $89.98$\\
\hline
$\beta/^\circ$ & $-89.10$ & $89.39$ \\
\hline
$m_{\beta\beta}/\text{meV}$ & 4.21 & 30.73\\
\hline
$J_{CP}$ & -0.035 & 0.033\\
\hline
\end{tabular}
\caption{ The maximum and minimum values of $m_1$, $m_2$, $m_3$, $\Sigma m_i$, $\alpha$, $\beta$, $m_{\beta\beta}$ and $J_{CP}$ for normal hierarchy.} 
\label{values of physical parameters NO}
\end{table*}

\newcolumntype{P}[1]{>{\centering\arraybackslash}p{#1}}
\begin{table*}
\centering
\begin{tabular} {c c c}
\hline
Parameters & Minimum Value & Maximum Value\\
\hline
\hline
$m_1 /\text{eV}$ & 0.0507 & 0.0526\\
\hline
$m_2 /\text{eV}$ & 0.0514 & 0.0533 \\
\hline
$m_3 /\text{eV}$ & 0.0128 & 0.016\\
\hline
$\sum{m_i} /\text{eV}$ &0.114 &0.12\\
\hline
$|\alpha|/^\circ$ & 75.05 & $89.98$\\
\hline
$|\beta|/^\circ$ & 67.87 & $89.97$ \\
\hline
$m_{\beta\beta}/\text{meV}$ & 49.54 & 51.80\\
\hline
$J_{CP}$ & $-1.64\times10^{-6}$ & -0.021\\
\hline
\end{tabular}
\caption{ The maximum and minimum values of $m_1$, $m_2$, $m_3$, $\Sigma m_i$, $\alpha$, $\beta$, $m_{\beta\beta}$ and $J_{CP}$ for inverted hierarchy.} 
\label{values of physical parameters IO}
\end{table*}

\begin{figure}
  \centering
    \subfigure[]{\includegraphics[width=0.31\textwidth]{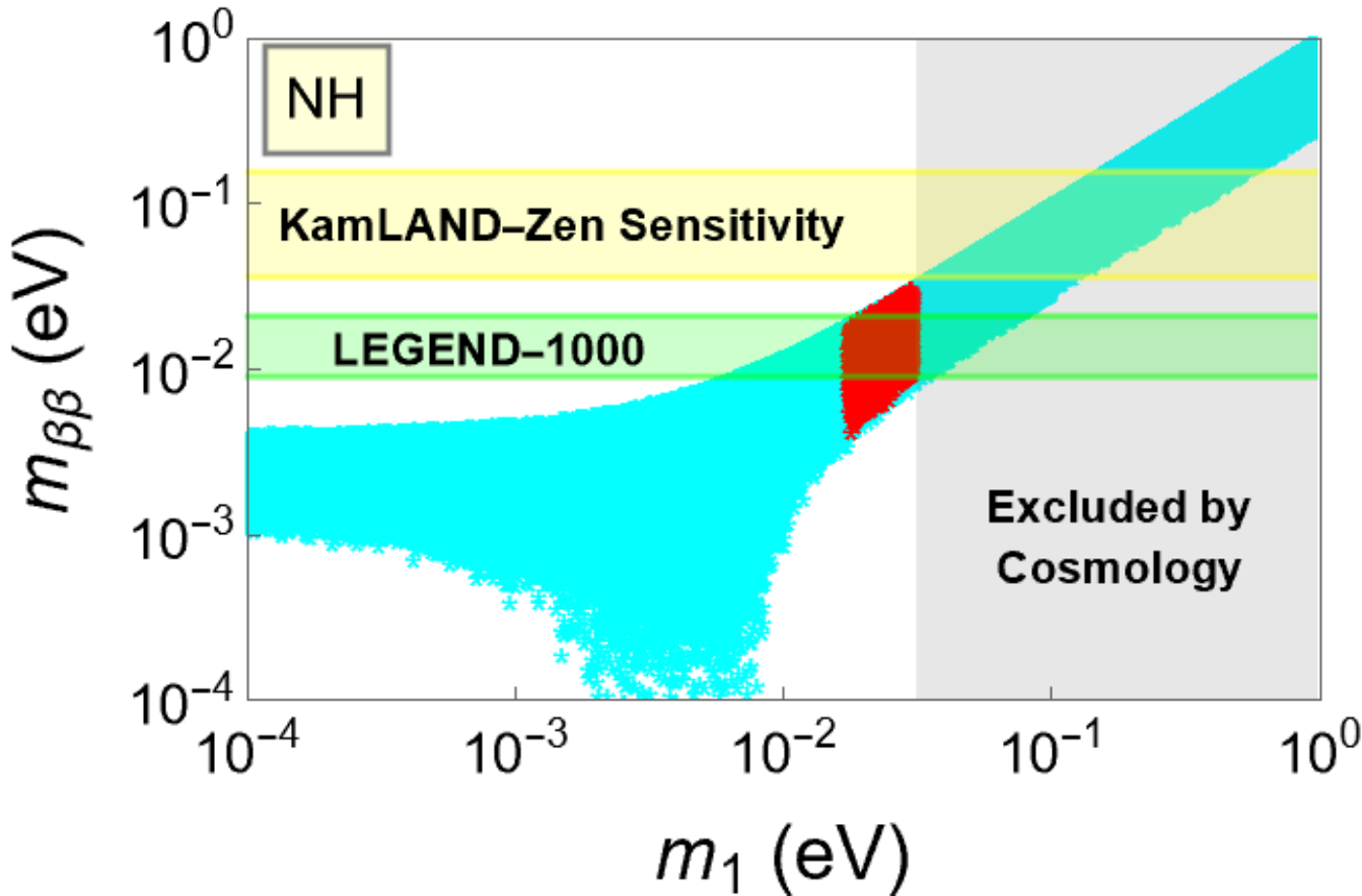}\label{fig:6a}} 
    \subfigure[]{\includegraphics[width=0.31\textwidth]{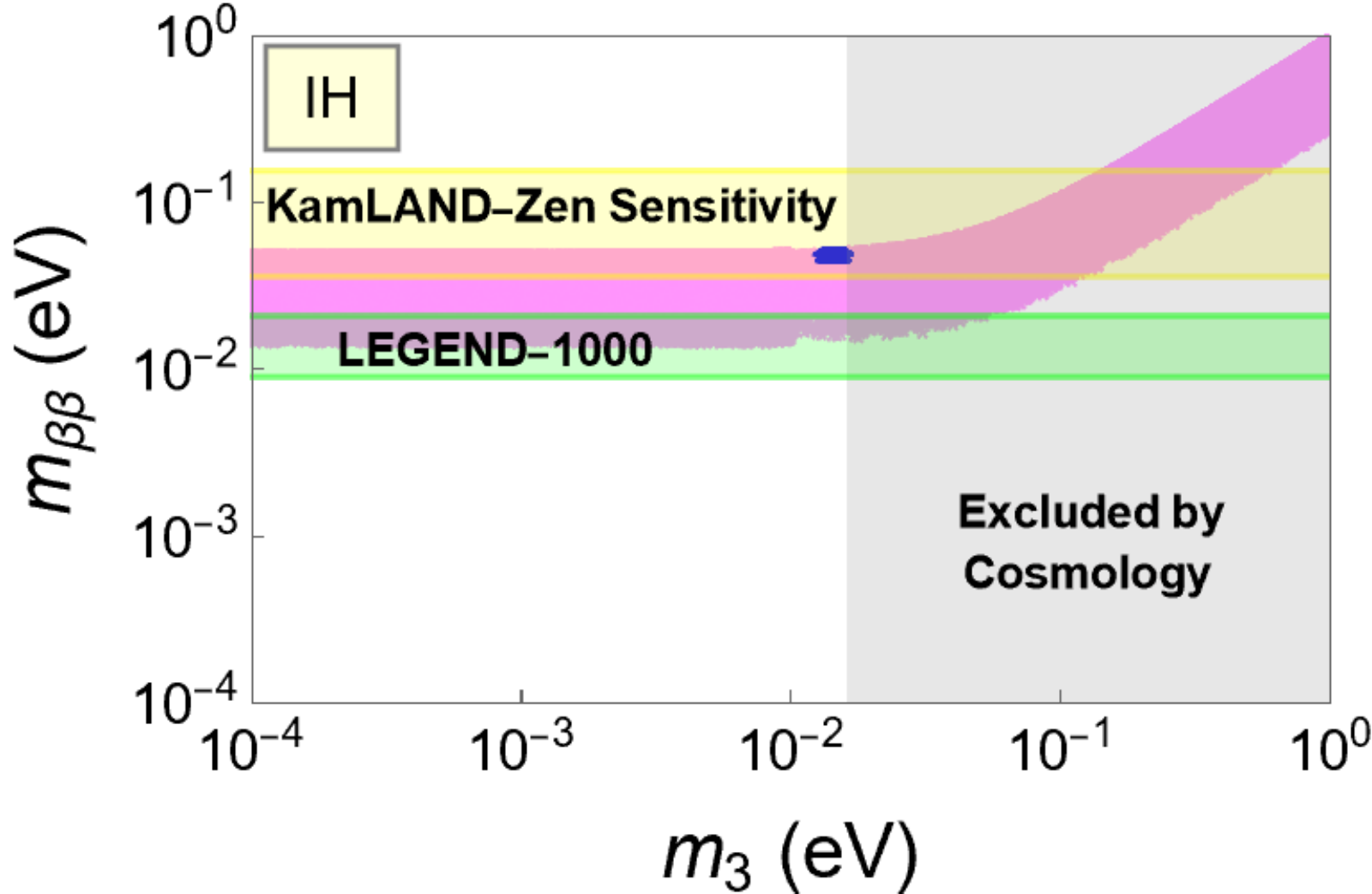}\label{fig:6b}}
   \caption{ The prediction of $m_{\beta\beta}$ for (a) Normal hierarchy, (b) Inverted hierarchy.}
\label{fig:mbb}
\end{figure}

We wish to highlight some interesting features of the proposed texture:

\begin{itemize}
\item For normal hierarchy, the texture is valid for all values of the mixing angles within the $3\sigma$ range. This is also true for inverted hierarchy, except for $\theta_{23}$, which lies in the upper octant and is constrained within a narrow region, with its minimum and maximum values approximately equal to $50^\circ$ and $52^\circ$ respectively\,(see Fig.\,\ref{fig:1b}).

\item It is interesting to note that the Dirac CP phase $\delta$ encounters a forbidden region under the proposed texture. For normal hierarchy, the maximum value of $\delta$ is approximately $310^\circ$ (see Fig.\,\ref{fig:1a}), whereas, the experiments give an upper bound of $\delta$ at $404^\circ$. On the contrary, for inverted hierarchy, the constraint on $\delta$ is much more stringent: $319.35^\circ<\delta<360^\circ$\,(see Fig.\,\ref{fig:1b}).

\item The proposed texture predicts sharp bounds on the Majorana phases as well. For normal hierarchy, $68^\circ<|\alpha|<90^\circ$ (see Fig.\,\ref{fig:1c}), whereas for inverted hierarchy, $75.05^\circ<|\alpha|<90^\circ$\,(see Fig.\,\ref{fig:1d}). For normal hierarchy, $-89^\circ<\beta<89.39^\circ$\,(see Fig.\,\ref{fig:1c}), whereas for inverted hierarchy, $68^\circ<|\beta|<90^\circ$\,(see Fig.\,\ref{fig:1d}).

\item For normal hierarchy, the numerical analysis is consistent with the mass spectra of $m_1$, $m_2$, and $m_3$ as follows: $0.0172\, \text{eV} < m_1 < 0.0304\,\text{eV}$, $0.01922\,\text{eV}< m_2 < 0.0316\,\text{eV}$, and $0.0524\,\text{eV}< m_3 < 0.059\,\text{eV}$ (see Fig.\,\ref{fig:2a}). Similarly, for inverted hierarchy,  $0.0507\,\text{eV} < m_1 < 0.0526\,\text{eV}$, $0.0514\,\text{eV} < m_2 < 0.0533\,\text{eV}$, and $0.0128\,\text{eV}< m_3 < 0.016\,\text{eV}$ (see Fig.\,\ref{fig:2b}). Here, we see that the spectra of $m_1$ and $m_2$ lie apart in the case of normal hierarchy, whereas for inverted hierarchy, there is an apparent overlapping. Nevertheless, this seems misleading as it hints at an exact degeneracy of $m_1$ and $m_2$ in the overlapping region. To address this confusion, we graphically analyze $m_2 - m_1$ in Fig.\,\ref{fig:2c} and verify that this difference never becomes exactly equal to zero. 

\end{itemize}

  For graphical visualization, we obtain the correlation plots for the sum of three neutrino mass eigenvalues\,(see Fig.\,\ref{fig:sum}). We know that the Jarlskog invariant ($J_{CP}$) \cite{Jarlskog:1985ht}, which determines the amount of the leptonic CP violation can be written from standard parametrization as shown below,
  
  \begin{equation}
  J_{CP}=s_{13} c_{13}^2 s_{12} c_{12} s_{23} c_{23} \sin\delta \nonumber.
  \end{equation}

From the proposed texture, we visualize the variation of $J_{CP}$ against Dirac CP phase $\delta$ in Fig.\,\ref{fig: jcp vs delta}. The Fig.\,(\ref{fig:texture parameters}) shows parameters space for the parameters Re$[G]$, Re$[J]$, and Im$[J]$ for both the hierarchies. The maximum and minimum values of the physical parameters are listed in Table\,(\ref{values of physical parameters NO}) and (\ref{values of physical parameters IO}) respectively. The parameter $m_{\beta \beta}$ which is the effective Majorana neutrino mass is an observational parameter. It can be expressed as shown in the following, \[m_{\beta \beta}=|\sum_{i=1}^{3}{U^2_{1i}m_i}|.\] where, $m_i$ signifies the three mass eigenvalues, $U_{11}$,$U_{12}$ and $U_{13}$ are the elements of the PMNS matrix. Several experiments have provided the upper bounds of $m_{\beta \beta}$: SuperNEMO(Se$^{82}$) as $67-131$ meV,  GERDA(Ge$^{76}$) as $104-228$ meV, EXO-200(Xe$^{136}$) as $111-477$ meV, CUORE(Te$^{130}$) as $75-350$ meV and KamLAND-Zen(Xe$^{136}$) as $61-165$ meV \cite{Ejiri:2020xmm, Agostini:2022zub, CUORE:2019yfd, GERDA:2019ivs, KamLAND-Zen:2016pfg, SuperNEMO:2021hqx, CUORE:2018ncg}. In this regard, the variation of the parameter $m_{\beta \beta}$ with respect to the lightest neutrino mass is visualized from the proposed texture\,(see Fig.\,(\ref{fig:mbb})). For the normal hierarchy, it is seen that the prediction of the parameter $m_{\beta \beta}$ aligns with the sensitivity of the future experiment LEGEND-1000\,\cite{LEGEND:2021bnm}. On the other hand, for inverted hierarchy, the prediction of $m_{\beta \beta}$ falls within the sensitivity of KamLAND-Zen experiment\,\cite{KamLAND-Zen:2016pfg}.

\section{The Model \label{section 4}}

To realize the texture shown in the earlier section, compatible with the proposed mixing scheme, we now consider a framework based on non-Abelian discrete flavour symmetry. As an example, we have considered $A_4$ discrete flavour symmetry, which is the smallest group having a three-dimensional irreducible representation where three SM lepton doublets can be accommodated and unified into the triplet representation. Here, the whole framework is embedded within $A_4 \times Z_{10} \times Z_2$ symmetry. The light neutrino mass is generated via the most general type-I+II seesaw mechanism\,\cite{Ma:2002nn}. Hence, in this regard, we extend the field content of the SM by adding three right-handed neutrinos\,$(\nu_{e_R},\nu_{\mu_R},\nu_{\tau_R})$ and two scalar triplets\,($\Delta_{1,2}$). To implement the flavour symmetry and appropriate texture of the mass matrix, we also include seven scalar singlets\,( $\Phi$, $\xi, \eta, \chi$, $\psi$, $\kappa$, $\varsigma$). The transformation properties of the field content associated with $A_4 \times Z_{10} \times Z_2$ symmetry are summarised in Table\,(\ref{Field Content of M2}).

\begin{table*}[h]
\centering
\begin{tabular}{P{1.5cm}P{0.5cm}P{2cm}P{2cm}P{0.6cm}P{0.6cm}P{0.6cm}P{0.6cm}P{0.5cm}P{0.5cm}P{0.5cm}P{0.5cm}P{0.5cm}P{0.5cm}} 
\hline
Fields & $D_{l_{L}}$ & $l_{R}$ & $\nu_{l_R}$ & $H$ & $\Delta_1$ & $\Delta_2$ & $\Phi$ & $\xi$ & $\eta$ &  $\chi$  & $\psi$ & $\kappa$ & $\varsigma$\\ 
\hline\hline
$SU(2)_{L}$ & 2 & 1 & 1 & 2 & 3 & 3 & 1 & 1 & 1 & 1  & 1 & 1 & 1\\
\hline
$U(1)_Y$ & -1 & $(-2,-2,-2)$ & $(0,0,0)$ & $1$ & -2 & -2 & $0$ & $0$ & $0$ & 0 & 0 & 0 & 0  \\
\hline
$A_4$ & 3 & $(1,1'',1')$ & $(1'',1,1^{'})$ & $1$ & 1 & 1 & $3$ & $1$ & $1''$ & $1$ & 3 & 3 & 1 \\
\hline
$Z_{10}$ & 0 & $(0,0,0)$ & $(0,4,8)$ & $0$ & 3 & 1 & $0$ & $2$ & $5$ & 6 & 7 & 9 & 5  \\
\hline
$Z_{2}$ & 1 & $(1,1,1)$ & $(1,1,1)$ & $-1$ & 1& 1 & $-1$ & $1$ & $1$ & 1 & 1 & 1 & 1  \\
\hline
\end{tabular}
\caption{ The transformation properties of various fields under $A_4 \times Z_{10} \times Z_2$.} 
\label{Field Content of M2}
\end{table*}

The $A_4 \times Z_{10} \times Z_2$ invariant Lagrangian can be constructed in the following manner,

\begin{eqnarray}
- \mathcal{L}_Y &=& \frac{y_e}{\Lambda} (\overline{D}_{l_L}\Phi)_1 H \,e_{R_1} + \frac{y_\mu}{\Lambda} (\overline{D}_{l_L} \Phi)_{1'} \,H \mu_{R_{1''}}+\,\frac{y_\tau}{\Lambda}\,(\overline{D}_{l_L} \Phi)_{1''} H \tau_{R_{1'}} + \frac{y_1}{\Lambda}\,(\overline{D}_{l_L} \Phi)_{1'}\nonumber\\&&\,\tilde{H} \nu_{e_{R_{1''}}} + \frac{y_2}{\Lambda^2} (\overline{D}_{l_L}\,\Phi)_{1} \,\tilde{H} \nu_{{\mu}_{R_{1}}} \chi+ \frac{y_3}{\Lambda^2}(\overline{D}_{l_L}\Phi)_{1''}\,\tilde{H}\nu_{{\tau}_{R_{1'}}} \xi+\,\frac{y_{a}}{2}\,\,(\overline{\nu^c}_{\mu_R}\,\nu_{\mu_R})_1\,\xi_1 \nonumber\\&&+ \frac{y_{b}}{2}\,[(\overline{\nu^c}_{e_{R}}\,\nu_{\tau_{R}})\,+ (\overline{\nu^c}_{\tau_{R}}\,\nu_{e_{R}})]_1\,{\xi_1}+\,\frac{y_{c}}{2\Lambda}\,(\overline{\nu^c}_{e_R}\,\nu_{e_R})_1\,\eta_{1''}\,\varsigma +\, \frac{y_t}{\Lambda}\,(\overline{D}_{l_L}\, D_{l_L}^c)_{3_S}\,\psi \Delta_1 \nonumber\\&&+\,\frac{y_s}{\Lambda} \,(\overline{D}_{l_L}\, D_{l_L}^c)_{3_S}\,\kappa \,\Delta_2\,+h.c.\nonumber\\
\label{Yukawa Lagrangian M2}
\end{eqnarray}

 The product rules under $A_4$ are given in \ref{C}. The details for the $A_4$ vacuum alignment have been discussed in \ref{b}.  
 Auxiliary symmetries always play an important role in flavor model building~\cite{Altarelli:2005yp}. In our framework, we have considered auxiliary symmetry groups $Z_{10}$ and $Z_2$  to restrict the next to leading order corrections. The details of $Z_{10}$ group can be found in Refs. \cite{CarcamoHernandez:2020udg, Dey:2024ctx, Dey:2023bfa}. Here $\Lambda$ is the cut-off scale of the theory. The charged lepton mass matrix under the choice of vacuum alignment $\langle \Phi \rangle_{\circ}=v_{\Phi}\,(1,0,0)^{T}$ and  $\langle H \rangle_{\circ}=v_{H}$ can be derived from Eq.(\ref{Yukawa Lagrangian M2}) as shown below,

\begin{equation}
 M_{l} = 
 \begin{bmatrix}
  \frac{y_e}{\Lambda} v_{\Phi}\,v_{H} &  0 &  0\\
 0 & \frac{y_\mu}{\Lambda} v_{\Phi}\,v_{H} & 0\\
 0 & 0 & \frac{y_\tau}{\Lambda} v_{\Phi}\,v_{H}\\
 \end{bmatrix},
 \label{chargrd lepton mass matrix}
\end{equation}

We choose the vacuum alignments $\langle\eta\rangle_{\circ}=v_{\eta}$, $\langle\chi\rangle_{\circ}=v_{\chi}$, $\langle\xi\rangle_{\circ}=v_{\xi}$, $\langle\varsigma\rangle_{\circ}=v_{\varsigma}$ to derive the Dirac neutrino mass matrix ($M_D$) and right handed neutrino mass matrix ($M_R$) as shown in the following,

\begin{eqnarray}
 M_{D} &=&
 \begin{bmatrix}
  0 &  \frac{y_2}{\Lambda^2}\,v_\Phi\,v_H v_\chi & 0\\
 \frac{y_1}{\Lambda} v_\Phi\,v_H & 0 & 0 \\
 0 & 0  & \frac{y_3}{\Lambda^2}\,v_\Phi\,v_H \,v_\xi\\
 \end{bmatrix},
 \label{Dirac Mass Matric M1}\\
 M_{R}&=&
 \begin{bmatrix}
\frac{y_c}{2\,\Lambda} \,v_\eta\,v_\varsigma & 0 & \frac{y_b}{2} v_\xi\\
 0 & \frac{y_a}{2} v_\xi & 0 \\
 \frac{y_b}{2} v_\xi & 0  & 0 \\ 
 \end{bmatrix},
 \label{M_R for M1}
\end{eqnarray}

The two matrices $M^1_{T_2}$ and $M^2_{T_2}$ from two type-II seesaw mechanisms can be derived from Eq.(\ref{Yukawa Lagrangian M2}) by considering the vacuum alignment$\langle\Delta_1\rangle_{\circ}=v_{\Delta_1}$, $\langle\Delta_2\rangle_{\circ}=v_{\Delta_2}$, $\langle\psi\rangle_{\circ}=v_{\psi}(0,1,-1)^{T}$ and $\langle\kappa\rangle_{\circ}=v_{\kappa}(0,1,-1)^{T}$ as shown in the following,

\begin{eqnarray}
 M^1_{T_2} &=&
 \begin{bmatrix}
  0 & \frac{1}{3\,\Lambda}\,y_t\,v_{\Delta_1}\,v_{\psi} & -\frac{1}{3\,\Lambda}\,y_t\,v_{\Delta_1}\,v_{\psi}\\
 \frac{1}{3\,\Lambda}\,y_t\,v_{\Delta_1}\,v_{\psi} & \frac{2}{3\,\Lambda}\,y_t\,v_{\Delta_1}\,v_{\psi} & 0 \\
-\frac{1}{3\,\Lambda}\,y_t\,v_{\Delta_1}\,v_{\psi} & 0  & -\frac{2}{3\,\Lambda}\,y_t\,v_{\Delta_1}\,v_{\psi}\\
 \end{bmatrix},
 \label{first type-2}\\
 M^2_{T_2}&=&
 \begin{bmatrix}
\frac{2}{3\,\Lambda}\,y_s\,v_{\kappa}\,v_{\Delta_2} & -\frac{1}{3\Lambda}\,y_s\,v_{\kappa}\,v_{\Delta_2} & 0\\
-\frac{1}{3\Lambda}\,y_s\,v_{\kappa}\,v_{\Delta_2} & 0 & -\frac{1}{3\Lambda}\,y_s\,v_{\kappa}\,v_{\Delta_2} \\
 0 & -\frac{1}{3\Lambda}\,y_s\,v_{\kappa}\,v_{\Delta_2} & \frac{2}{3\,\Lambda}\,y_s\,v_{\kappa}\,v_{\Delta_2}\\
 \end{bmatrix}.
 \label{second type-2}
\end{eqnarray}
 The effective neutrino mass matrix $M_{\nu}$ is constructed after taking the contributions from type-I ($-\,M_{D}M^{-1}_{R}M^{T}_{D}$) and two type-II seesaw mechanisms as shown below,

\begin{eqnarray}
M_\nu &=& \begin{bmatrix}
A &  B  &  F \\
B &  -2F  &  G \\
F & G &  J \\
\end{bmatrix},
\end{eqnarray}

where, $A$, $B$, $F$, $G$, $J$ are purely complex parameters. The said texture parameters can be expressed in terms of the model parameters as shown in the following,

\begin{eqnarray}
A &=& \frac{2\,y_s\,v_{\Delta_2}\,v_{\kappa }}{3\,\Lambda }-\frac{2\,y_2^2\, v_H^2\,v_{\chi }^2\,v_{\phi }^2}{\Lambda^4\,y_a\,v_{\xi }},\\
B &=& \frac{y_t\,v_{\Delta_1}\,v_{\psi}}{3\,\Lambda }-\frac{y_s\,v_{\Delta_2}\,v_{\kappa}}{3\,\Lambda},\\
F &=& -\frac{y_t\,v_{\Delta_1}\,v_{\psi }}{3\,\Lambda},\\
G &=& -\frac{2\,y_1\,y_3\,v_H^2\,v_{\phi }^2}{\Lambda^3\,y_b}-\frac{y_s\,v_{\Delta_2}\,v_{\kappa}}{3\,\Lambda},\\
J &=& \frac{2\,y_3^2\,y_c\,v_H^2\,v_{\zeta}\,v_{\eta }\,v_{\phi }^2}{\Lambda^5\,y_b^2}+\frac{2 \,y_s\,v_{\Delta_2}\,v_{\kappa }}{3 \,\Lambda}-\frac{2\,y_t\,v_{\Delta_1}\,v_{\psi}}{3\,\Lambda }.
\end{eqnarray}

It is important to mention that it is very difficult to predict the Yukawa couplings individually. However, the present work predicts the combinations of the model parameters $D_1=\frac{y_t v_{\Delta_1} v_{\psi }}{\Lambda }$, $D_2=\frac{y_s v_{\Delta_2} v_{\kappa }}{\Lambda }$, $D_3=\frac{\Lambda ^4 y_a v_{\xi }}{2 y_2^2 v_H^2 v_{\chi }^2 v_{\phi }^2}$, $D_4=\frac{\Lambda ^3 y_b}{2 y_1 y_3 v_H^2 v_{\phi }^2}$ and $D_5=\frac{y_c \Lambda v_{\zeta } v_{\eta }}{y_1^2 v_H^2 v_{\phi }^2}$. These combinations can be expressed in terms of the texture parameters $A$, $B$, $F$, $G$ and $J$ as shown below,

\begin{eqnarray}
D_1 &=& -3\,F,\\
D_2 &=& -3 (B + F),\\
D_3 &=& -\frac{2}{A+2 B+2 F},\\
D_4 &=& \frac{2}{B+F-G},\\
D_5 &=& \frac{2 (2 B+H)}{(B+F-G)^2}.
\end{eqnarray}

\begin{table}
\centering
\begin{tabular}{ccc}
\hline
Model Parameter & Minimum Value & Maximum Value\\
\hline
\hline
$|D_1|/\,\text{eV}$ & 0.0013 & 0.0884\\
\hline
$|D_2|/\,\text{eV}$ & 0.0072 & 0.0713 \\
\hline
$|D_3|^{-1}/\,\text{eV}$ & 0.00007 & 0.0350\\
\hline
$|D_4|^{-1}/\,\text{eV}$ & 0.0021 & 0.0320\\
\hline
$|D_5|^{-1}/\,\text{eV}$ & 0.00009 & 0.3416\\ 
\hline
$Arg[D_1]/^\circ$ &-179.96 &179.99\\
\hline
$Arg[D_2]/^\circ$ &-179.96 & 179.97\\
\hline
$Arg[D_3]/^\circ$ &-179.98 & 179.63\\
\hline
$Arg[D_4]/^\circ$ &-179.99 & 179.98\\
\hline
$Arg[D_5]/^\circ$ &-165.86 & 169.90\\
\hline
\end{tabular}
\caption{Allowed values of the model parameters in the light of normal hierarchy compatible with 3$\sigma$ range of neutrino oscillation data\cite{Gonzalez-Garcia:2021dve}}.
\label{values of the model parameters NO}
\end{table}

\begin{figure}
  \centering
    \subfigure[]{\includegraphics[width=0.3\textwidth]{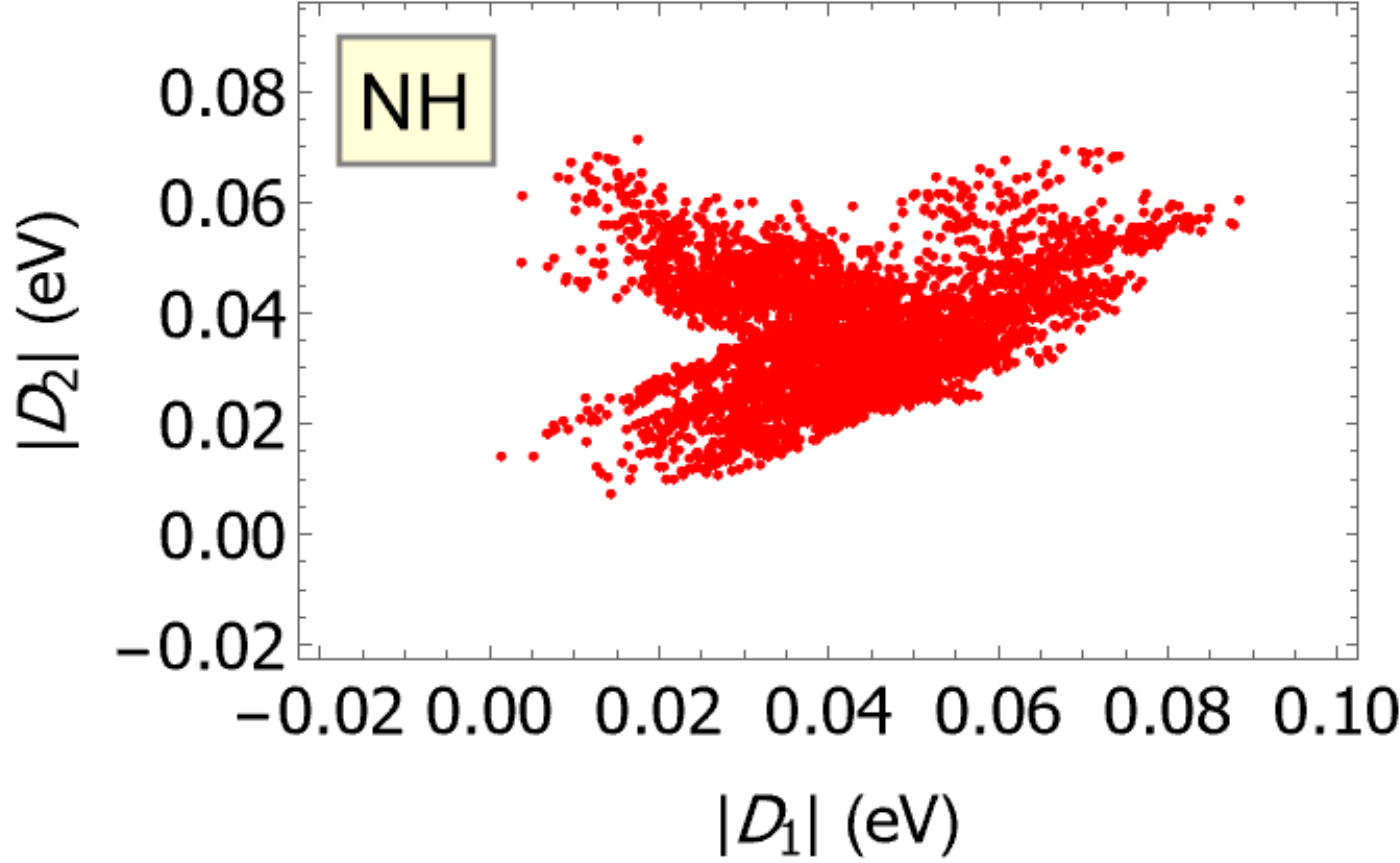}\label{fig:7a}} 
    \subfigure[]{\includegraphics[width=0.3\textwidth]{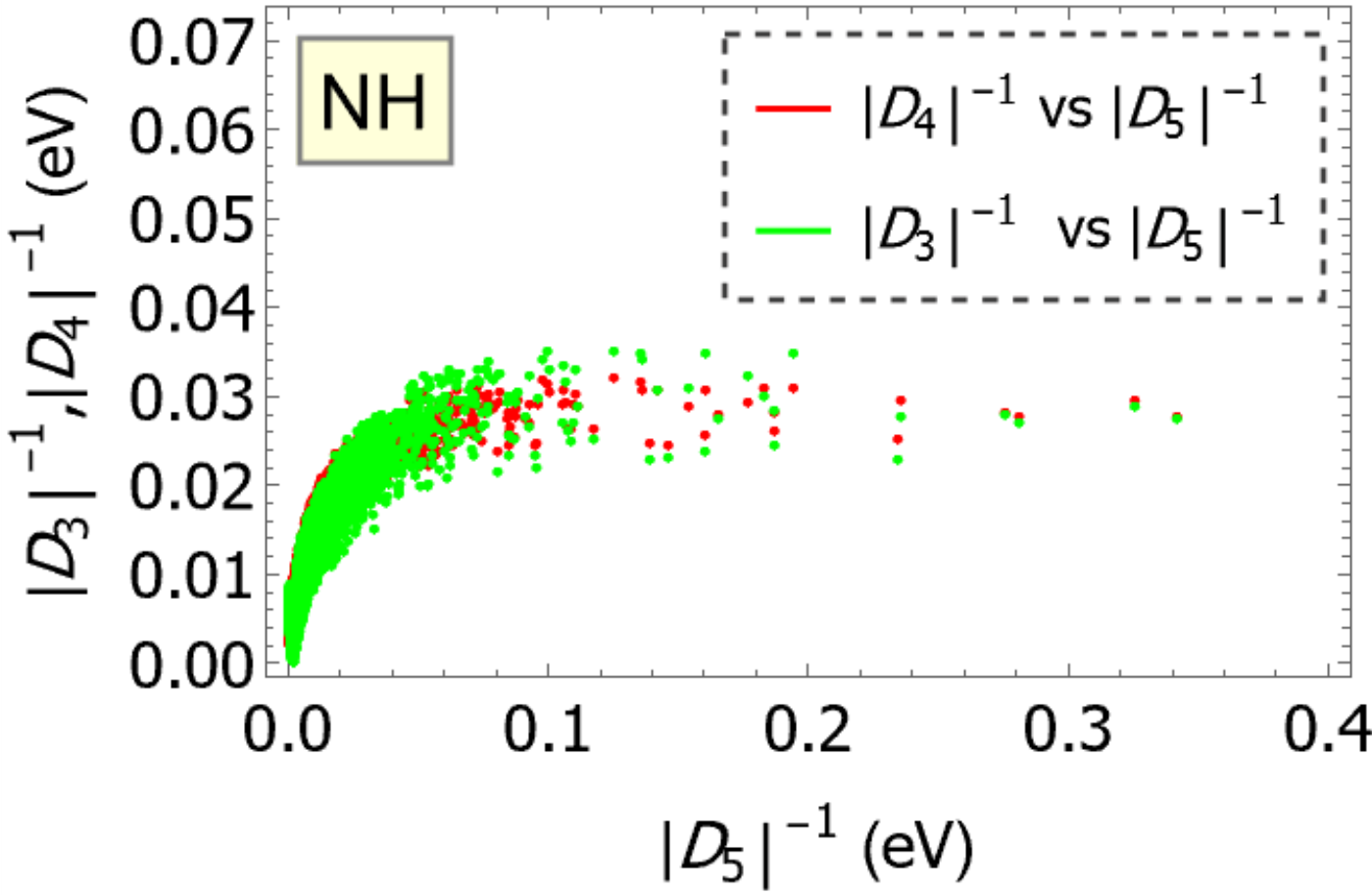}\label{fig:7b}}
    \subfigure[]{\includegraphics[width=0.3\textwidth]{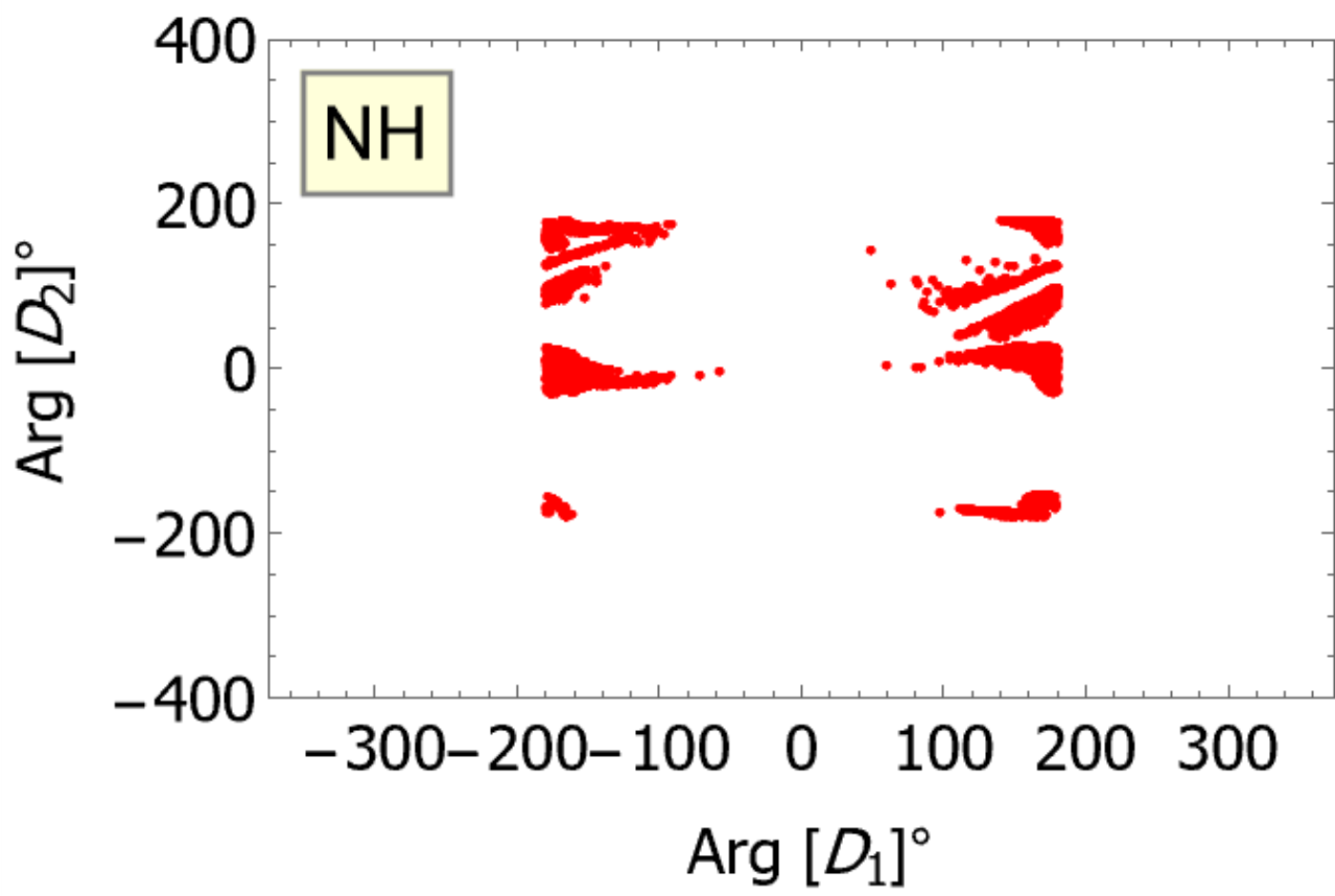}\label{fig:7c}}
     \subfigure[]{\includegraphics[width=0.3\textwidth]{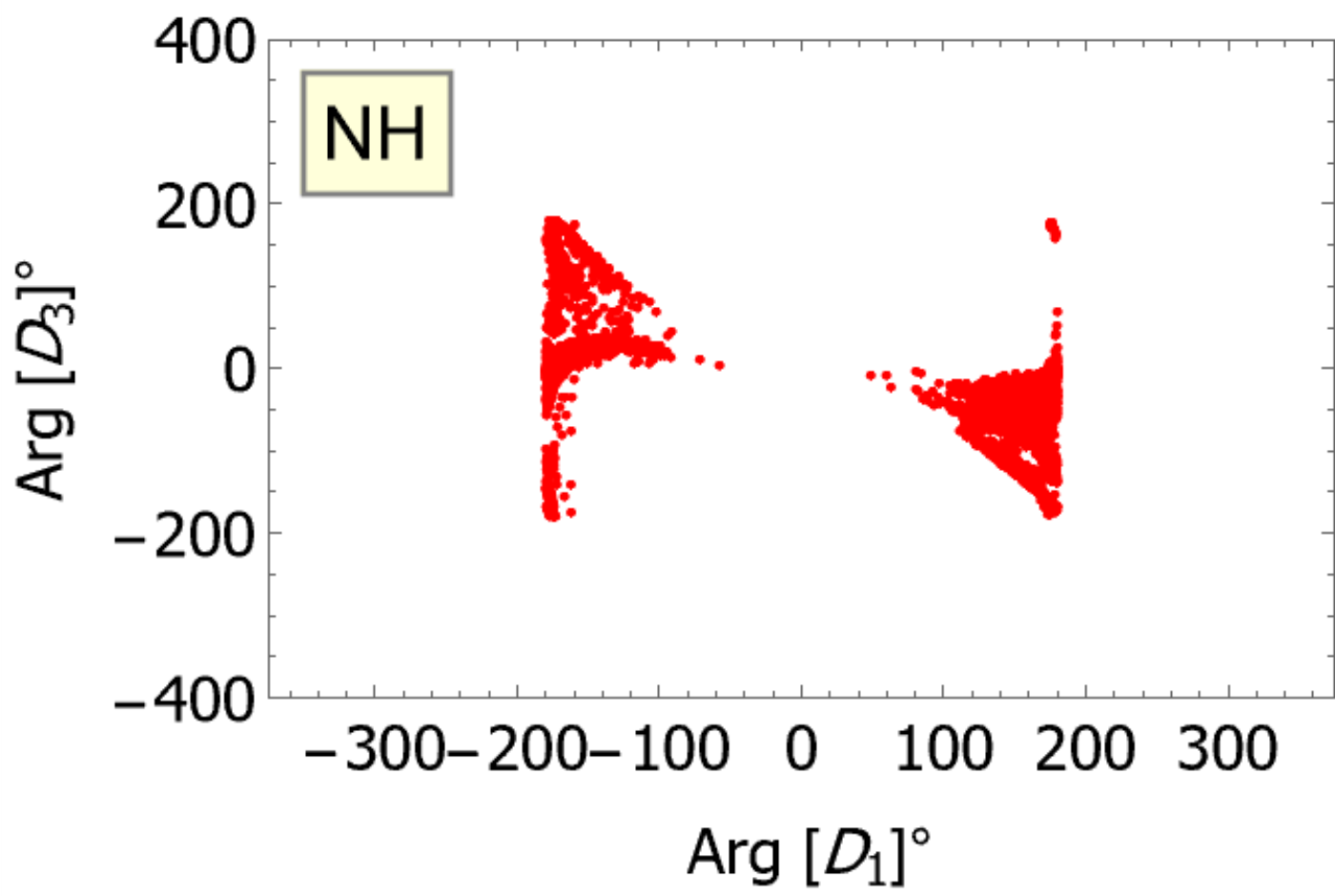}\label{fig:7d}} 
    \subfigure[]{\includegraphics[width=0.3\textwidth]{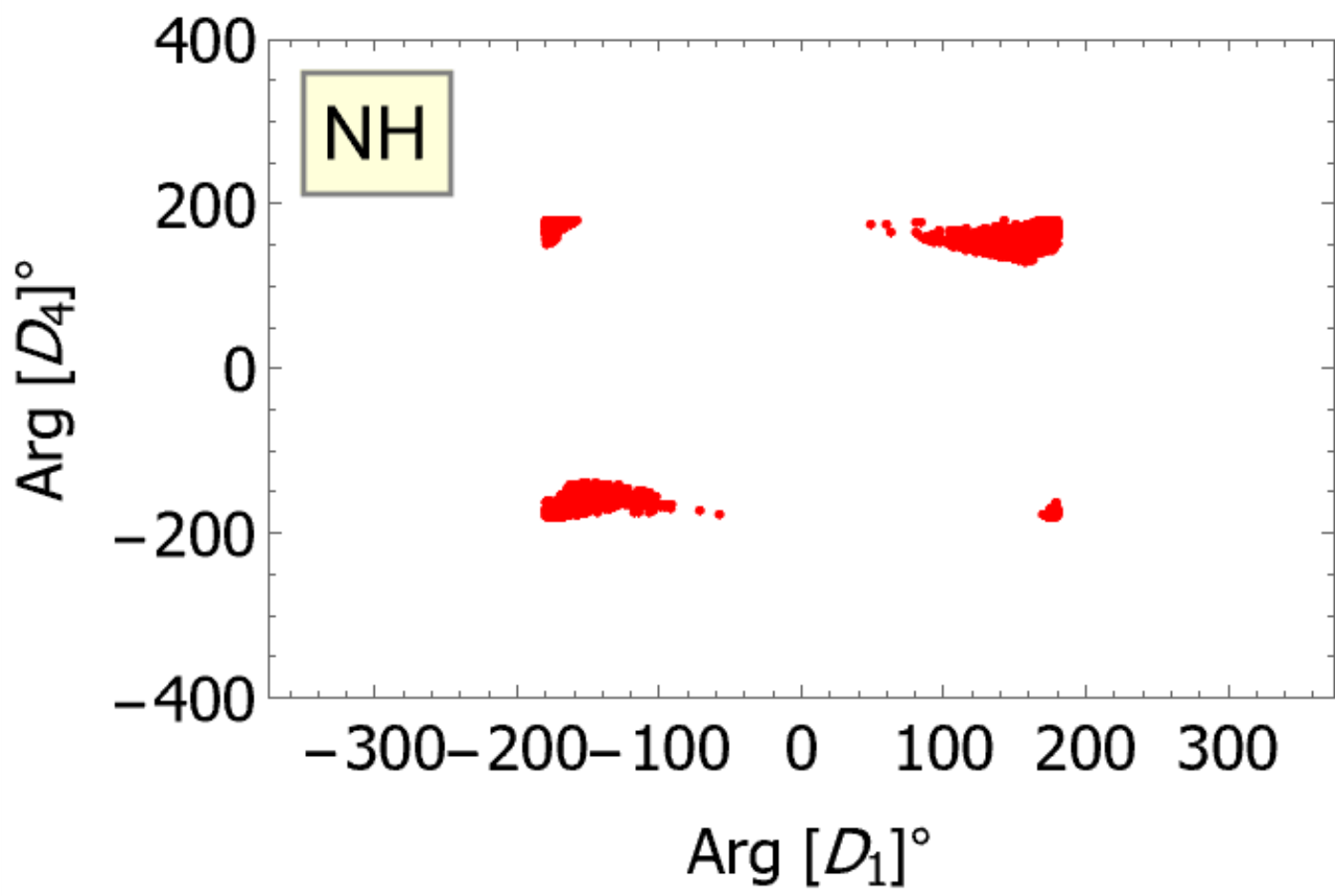}\label{fig:7e}}
    \subfigure[]{\includegraphics[width=0.3\textwidth]{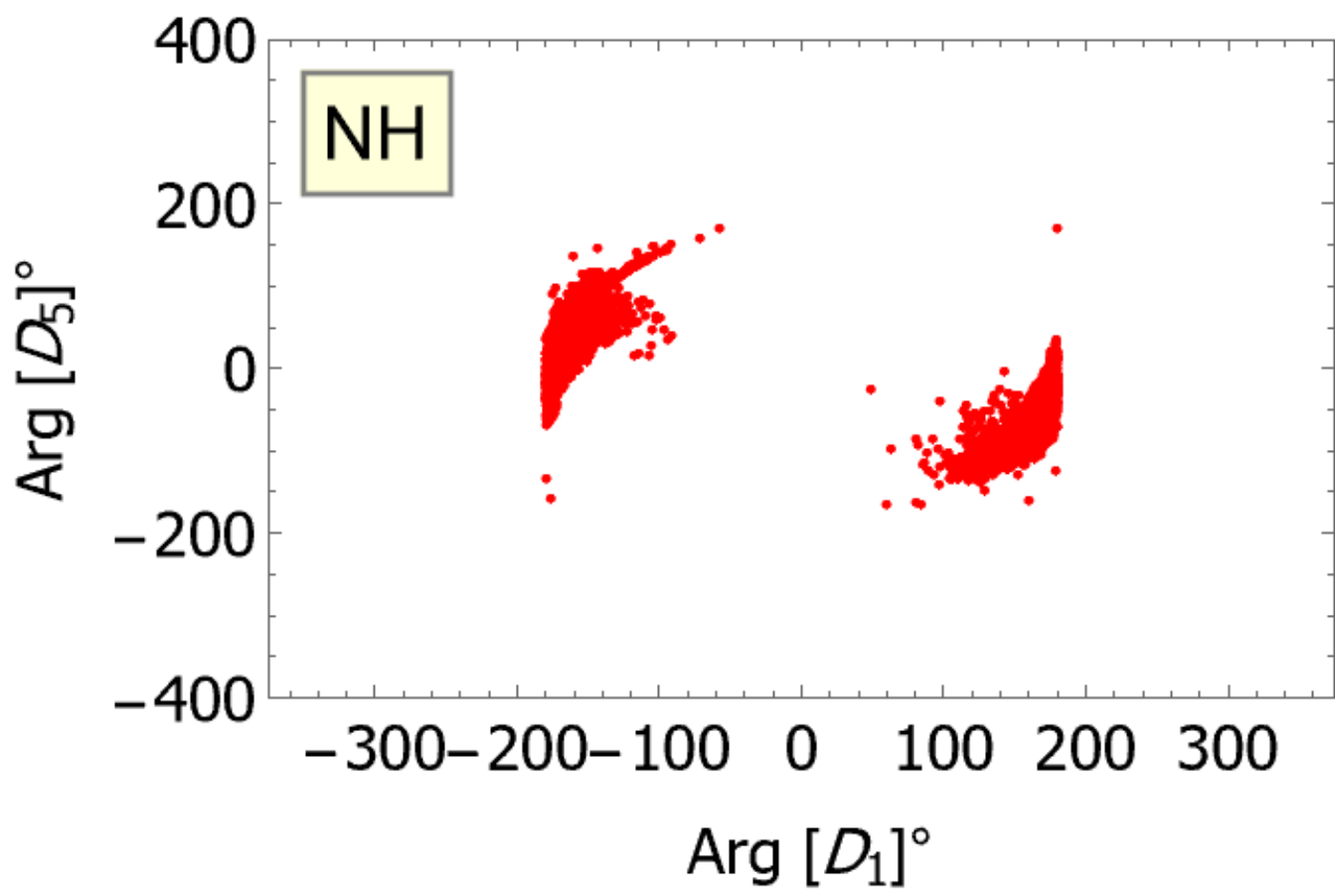}\label{fig:7f}}  
   \caption{ The correlation plots between (a) $|D_1|$ and $|D_2|$, (b) $|D_5|^{-1}$ vs $|D_3|^{-1}$ and $|D_4|^{-1}$, (c) Arg\,$[D_1]$ vs Arg\,$[D_2]$, (d) Arg\,$[D_1]$ vs Arg\,$[D_3]$, (e) Arg\,$[D_1]$ vs Arg\,$[D_4]$, (f) Arg\,$[D_1]$ vs Arg\,$[D_5]$ for normal hierarchy.}
\label{fig:model parameters NO}
\end{figure}

\begin{figure}
  \centering
    \subfigure[]{\includegraphics[width=0.3\textwidth]{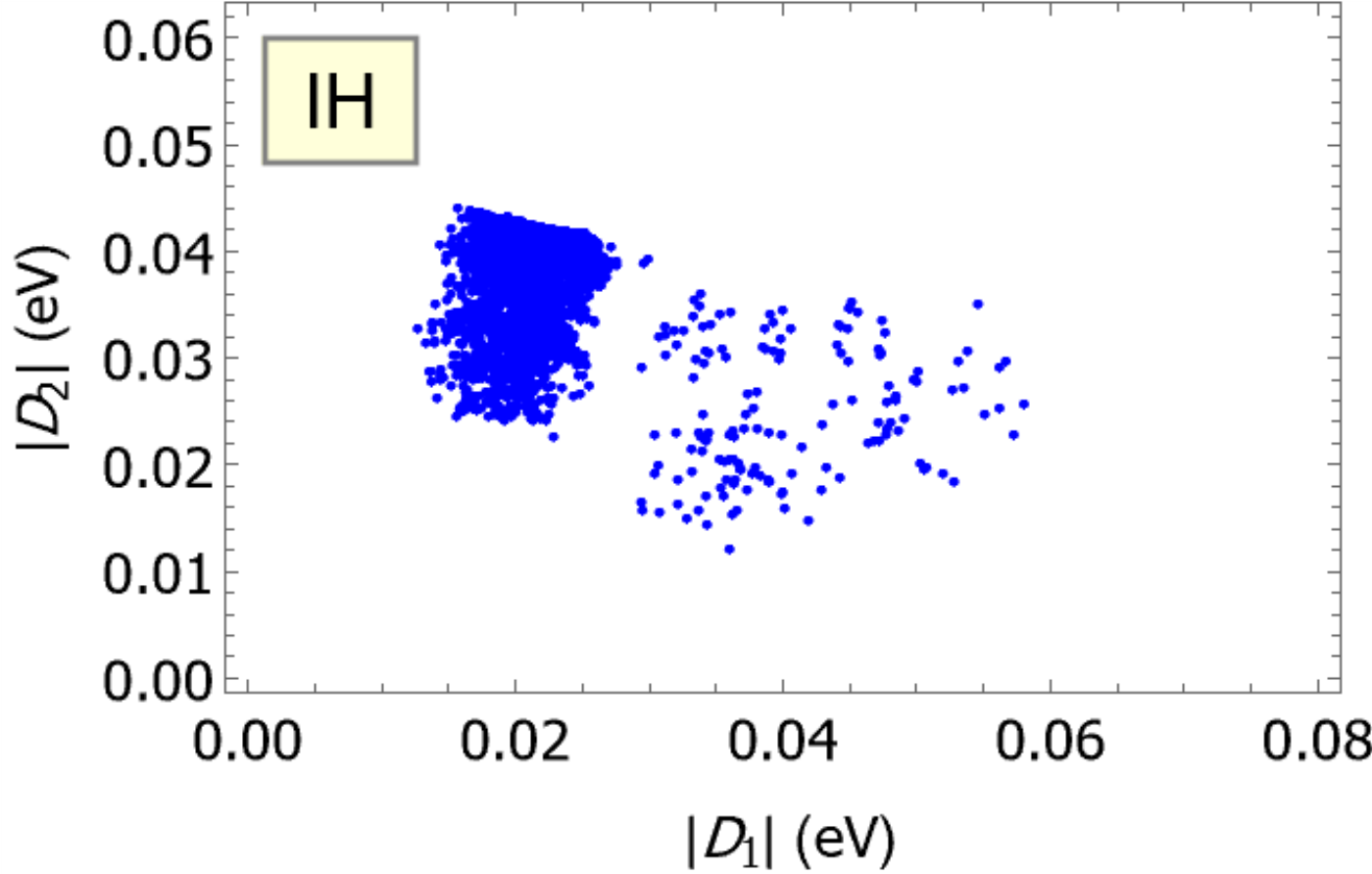}\label{fig:8a}} 
    \subfigure[]{\includegraphics[width=0.3\textwidth]{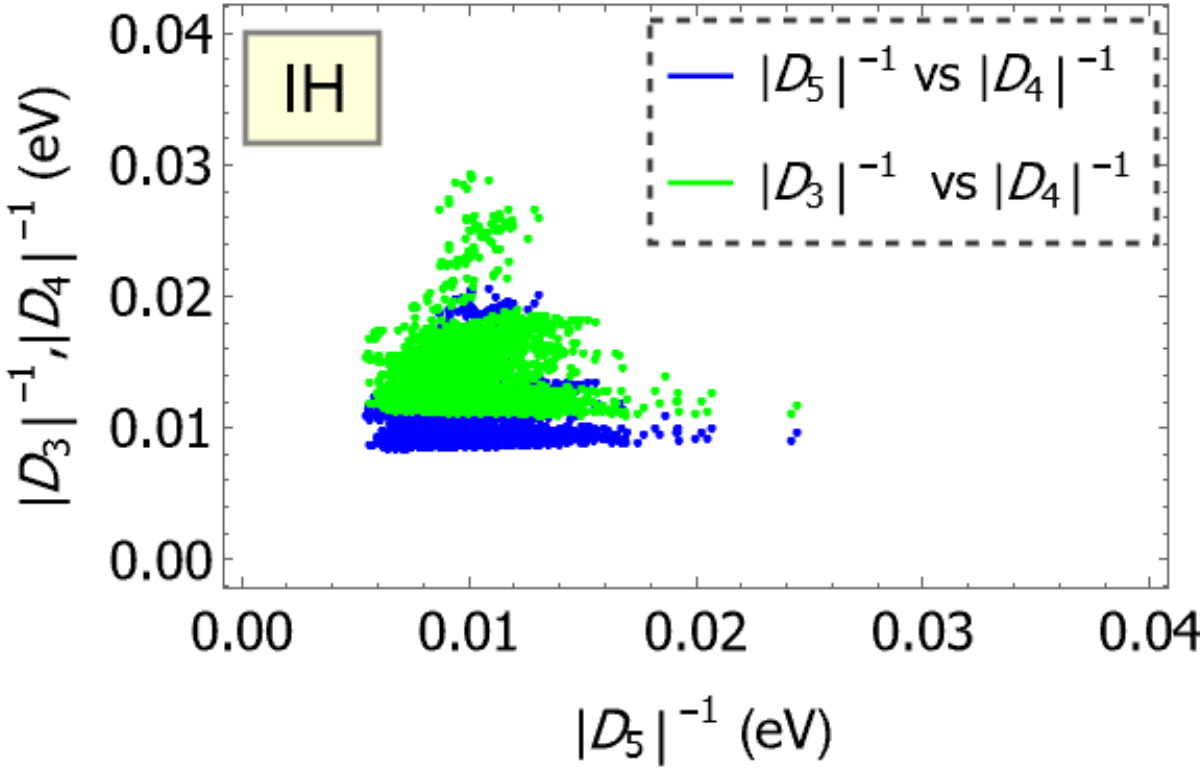}\label{fig:8b}}
    \subfigure[]{\includegraphics[width=0.3\textwidth]{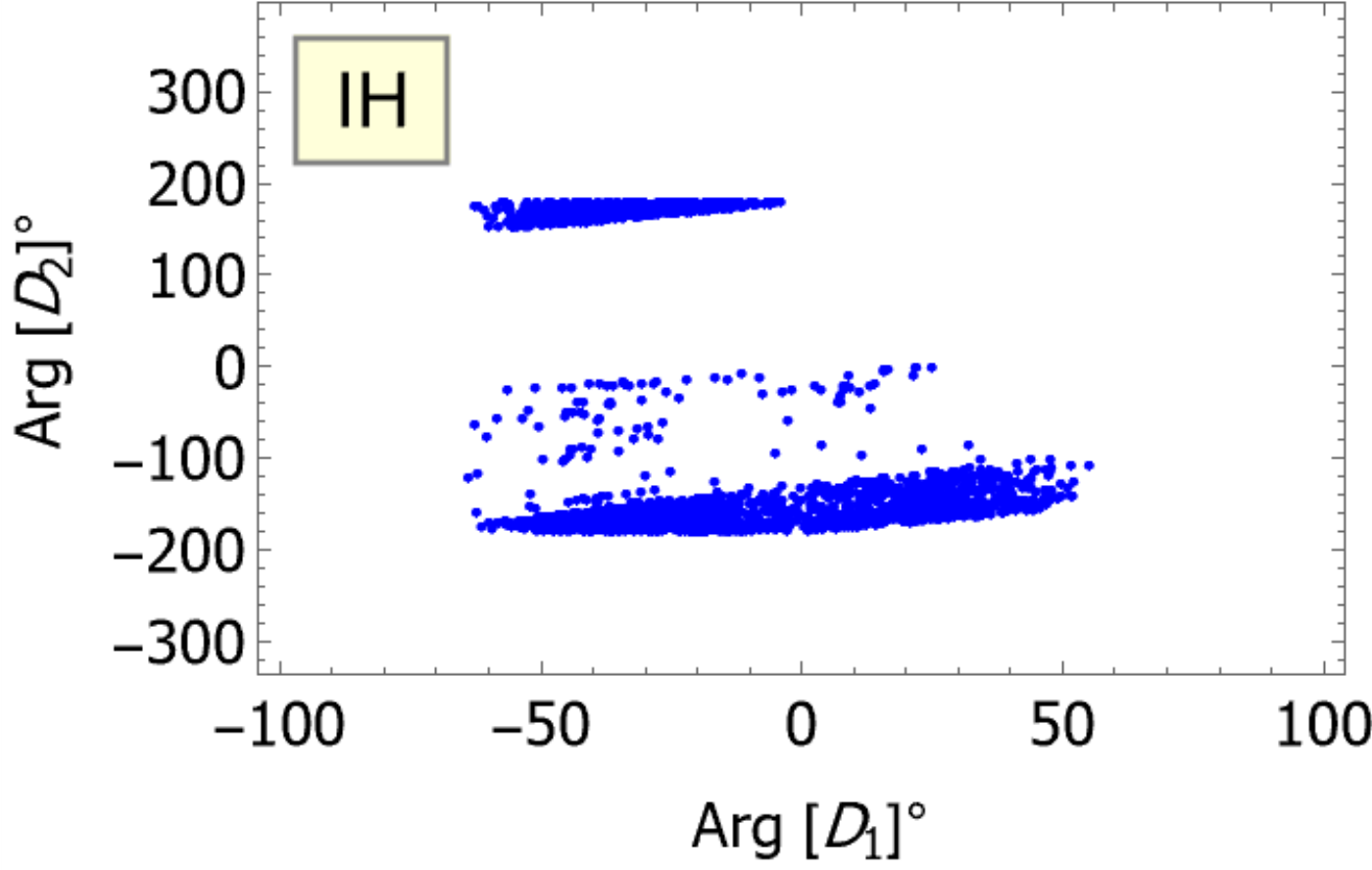}\label{fig:8c}}
     \subfigure[]{\includegraphics[width=0.3\textwidth]{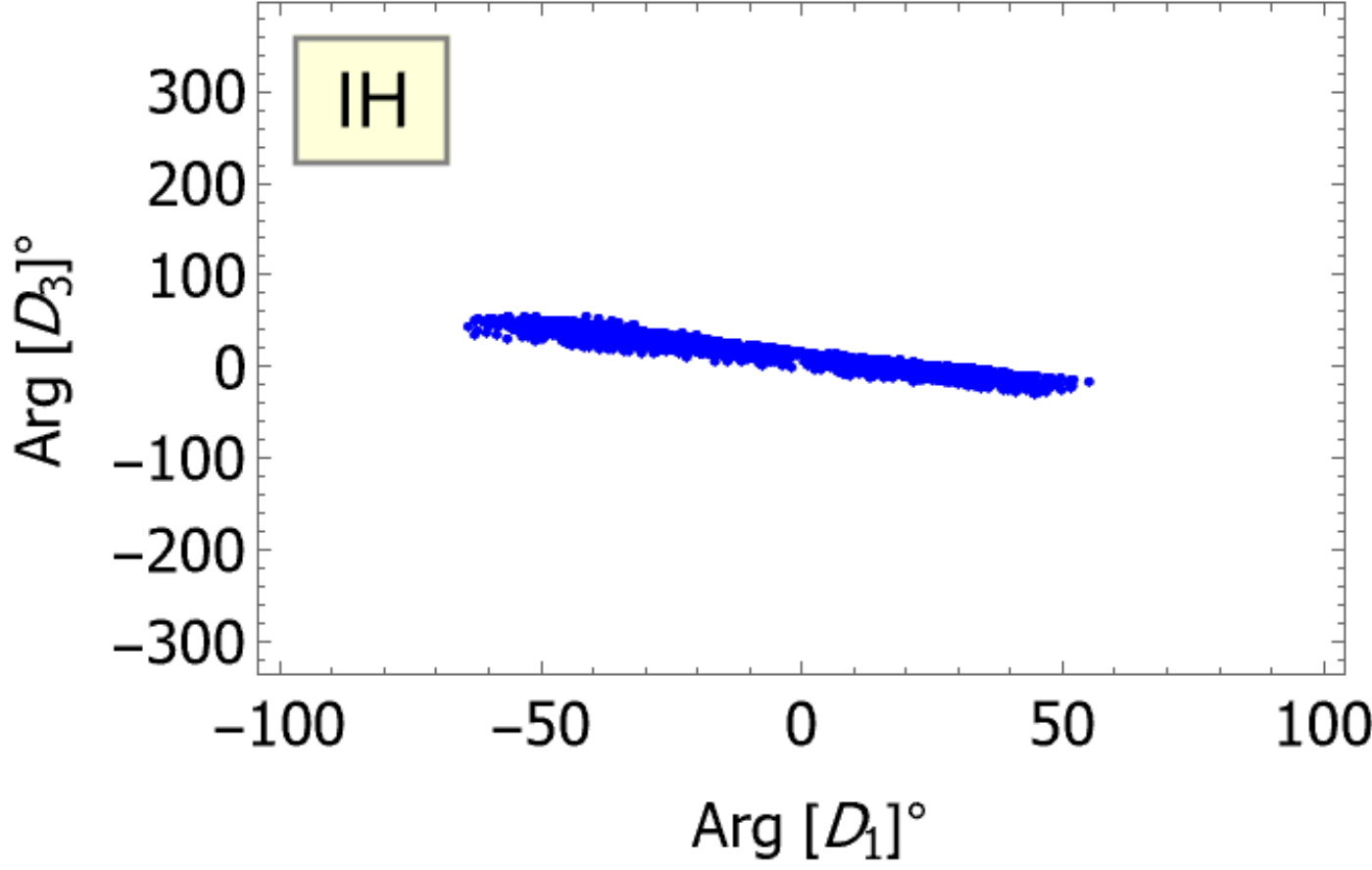}\label{fig:8d}} 
    \subfigure[]{\includegraphics[width=0.3\textwidth]{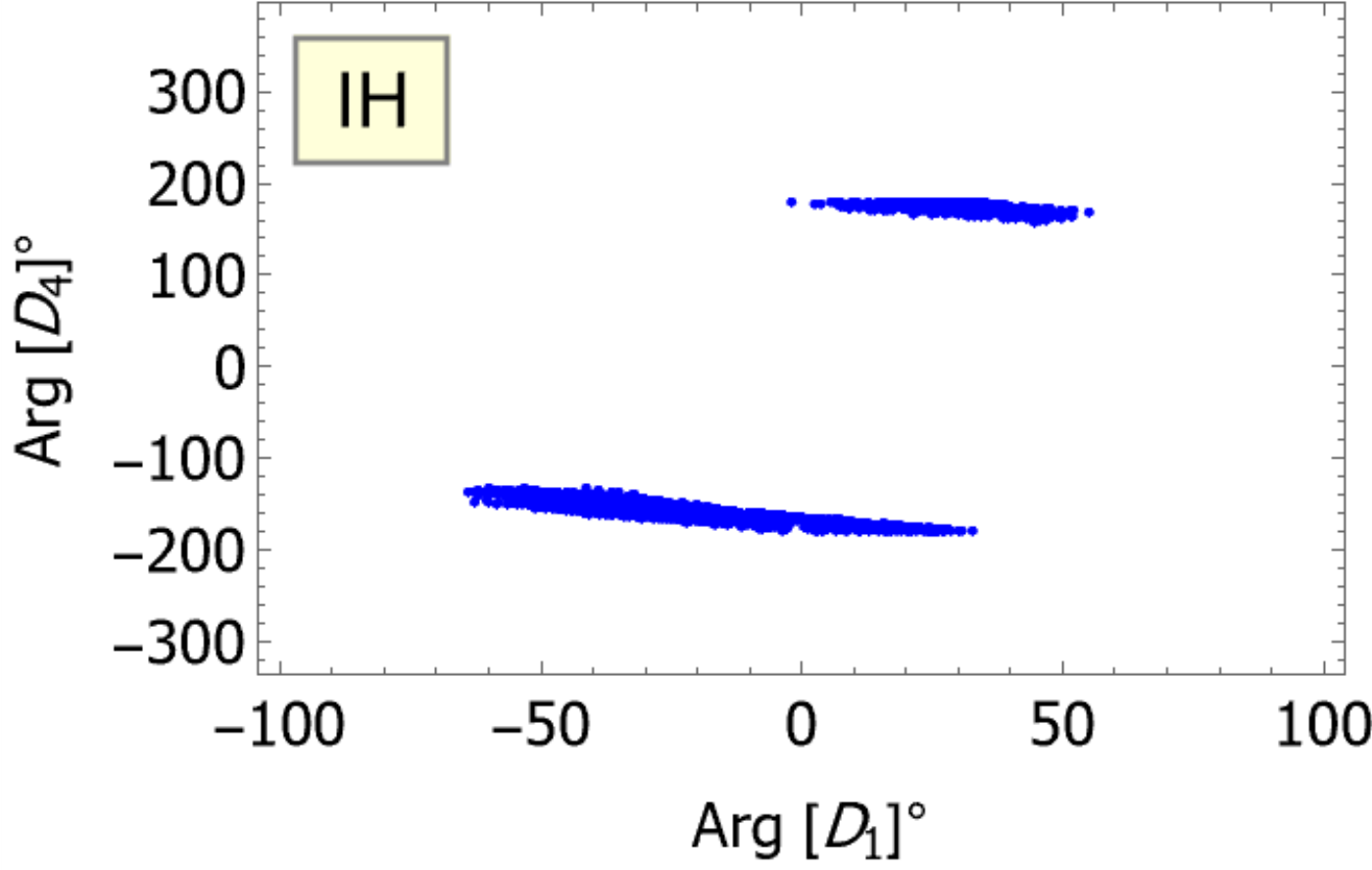}\label{fig:8e}}
    \subfigure[]{\includegraphics[width=0.3\textwidth]{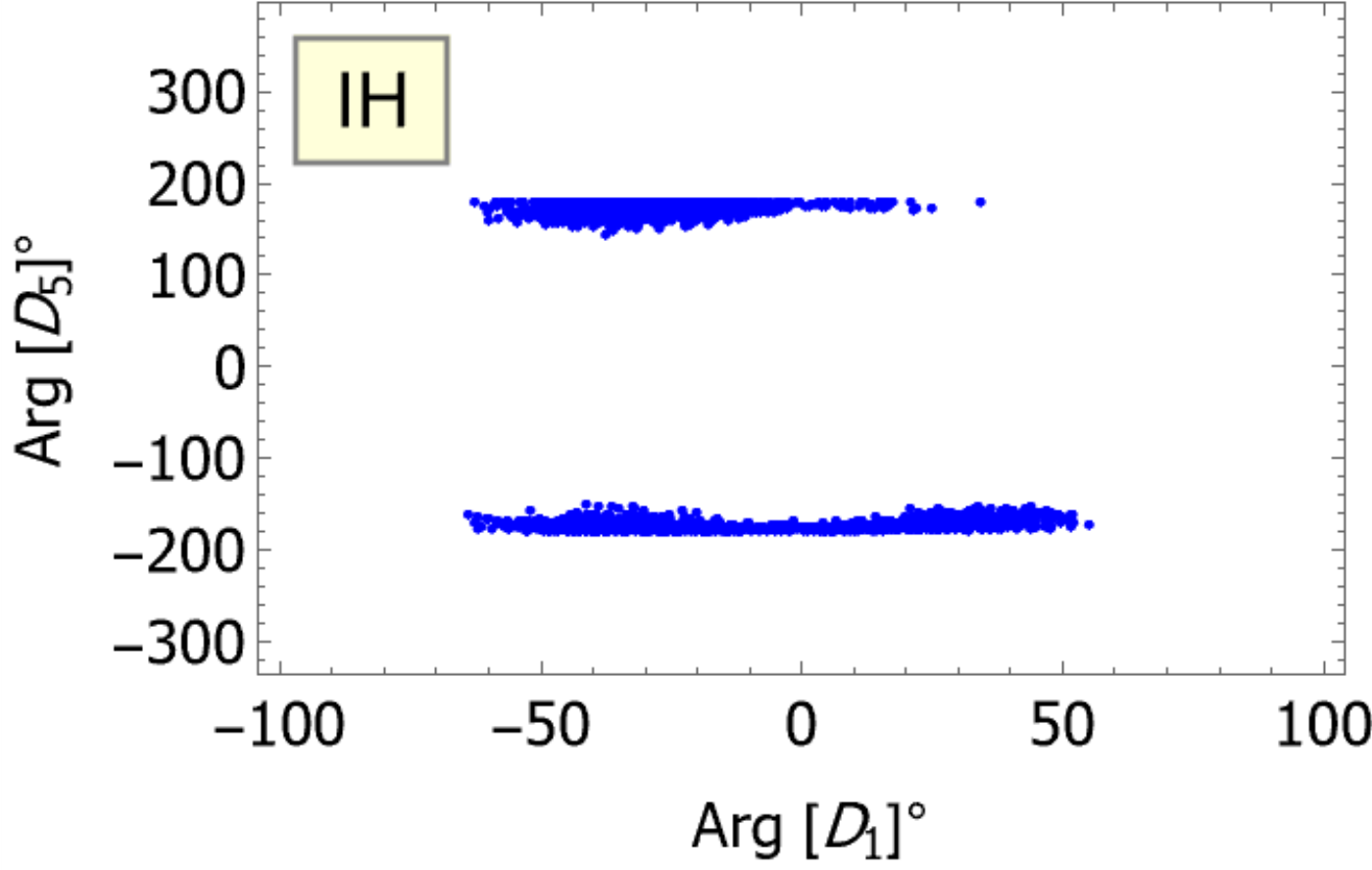}\label{fig:8f}}  
   \caption{ The correlation plots between (a) $|D_1|$ and $|D_2|$, (b) $|D_5|^{-1}$ vs $|D_3|^{-1}$ and $|D_4|^{-1}$, (c) Arg\,$[D_1]$ vs Arg\,$[D_2]$, (d) Arg\,$[D_1]$ vs Arg\,$[D_3]$, (e) Arg\,$[D_1]$ vs Arg\,$[D_4]$, (f) Arg\,$[D_1]$ vs Arg\,$[D_5]$  for inverted hierarchy.}
\label{fig:model parameters IO}
\end{figure}

\begin{table}
\centering
\begin{tabular}{ccc}
\hline
Model Parameter & Minimum Value & Maximum Value\\
\hline
\hline
$|D_1|/\,\text{eV}$ & 0.0126 & 0.0579\\
\hline
$|D_2|/\,\text{eV}$ & 0.0119 & 0.0441 \\
\hline
$|D_3|^{-1}/\,\text{eV}$ & 0.0108 & 0.0292\\
\hline
$|D_4|^{-1}/\,\text{eV}$ & 0.0083 & 0.0205\\
\hline
$|D_5|^{-1}/\,\text{eV}$ & 0.0054 & 0.0244\\ 
\hline
$Arg[D_1]/^\circ$ &-63.94 & 55.02\\
\hline
$Arg[D_2]/^\circ$ &-179.98 & 179.99\\
\hline
$Arg[D_3]/^\circ$ &-30.49 & 55.30\\
\hline
$Arg[D_4]/^\circ$ & -179.95 & 179.93\\
\hline
$Arg[D_5]/^\circ$ &-179.99 & 179.99\\
\hline
\end{tabular}
\caption{Allowed values of the model parameters in the light of inverted hierarchy compatible with 3$\sigma$ range of neutrino oscillation data\cite{Gonzalez-Garcia:2021dve}}.
\label{values of the model parameters IO}
\end{table}

For graphical representation, we generate correlation plots to visualize how the absolute values and arguments of the combinations of the model parameters vary. It is evident from Figs.\,(\ref{fig:model parameters NO}) and (\ref{fig:model parameters IO}) that the studied parameters are constrained to certain bounds for both normal and inverted hierarchies. The maximum and minimum values of the model parameters for both hierarchies, considering the 3$\sigma$ range of neutrino oscillation data, are listed in Tables (\ref{values of the model parameters NO}) and (\ref{values of the model parameters IO}) respectively\,~\cite{Gonzalez-Garcia:2021dve}.

Deriving the proposed model-independent neutrino mass matrix texture as shown in Eq. (\ref{eq:mass-mat-textute}) from the symmetry framework has some inherent challenges. Moreover, it's crucial to ensure the independence of parameters $A, B, F, G$, and $J$ from each other. This is achieved by including contributions from both type-I and type-II contributions, leading to an increase in the field content. This is a necessary step to achieve the high productivity of the framework. To obtain the texture in its exact form, we introduce the field $\varsigma$, to cut short certain non-leading terms in the Yukawa Lagrangian. Similarly, we need to cut short specific elements within the right-handed Majorana neutrino mass matrix which requires two additional scalar fields $\zeta$ and $\chi$. If $\Delta_1$ and $\Delta_2$ were chosen as $A_4$ triplets, the field content could have been reduced. However, in that case, certain components within $\Delta_i$ will not get vevs and consequently, there remains the finite possibility that these components will contribute towards the neutrino mass via radiative correction. To avoid this, we need to treat $\Delta_1$ and $\Delta_2$ as singlets under $A_4$ and introduce the $A_4$ scalar triplet fields $\psi$ and $\kappa$ to restore the invariance of the Lagrangian.

\section{Summary and Discussion \label{section 5}}

In the present work, we propose a neutrino mass matrix texture highlighting a suitable correlation $m_{22} = -2m_{13}$ among its elements. We run a phenomenological study in light of both normal and inverted hierarchies to study the viability of the proposed texture. In this regard, we observe that the texture imposes specific bounds on some of the observational parameters. Notably, the Dirac CP phase $\delta$ gets constrained when the analysis is performed for both hierarchies. Similarly, the atmospheric mixing angle $\theta_{23}$ lies in the upper octant for the inverted hierarchy. Additionally, the texture constrains two Majorana phases. The mass spectra for three neutrino mass eigenvalues ensure that $\sum m_i \leq 0.12$ eV,\cite{Planck:2018vyg}, and two mass squared differences $\Delta m_{21}^2$ and $\Delta m_{31}^2$ are consistent with experimental data,\cite{Gonzalez-Garcia:2021dve}. The effective mass parameter $m_{\beta\beta}$, appearing in neutrinoless double beta decay, is also predicted. For the normal hierarchy, it is seen that the prediction aligns with the sensitivity of future experiments such as LEGEND-1000,\cite{LEGEND:2021bnm}. For the inverted hierarchy, the prediction of $m_{\beta\beta}$ falls within the sensitivity of the KamLAND-Zen experiment,\cite{KamLAND-Zen:2016pfg}. We demonstrate that the texture can be derived from the interplay of general type-I+II seesaw mechanisms within the framework of the $A_4 \times Z_{10} \times Z_2$ group. As a future prospect, we aim to study the baryon asymmetry of the Universe and lepton flavour violating decays in the context of the proposed texture. The detailed analysis in this regard is left for future work.

\section{Acknowledgement} 

The research work of PC is supported by Innovation in Science Pursuit for Inspired Research (INSPIRE), Department of Science and Technology, Government of India, New Delhi, vide grant No. IF190651. MD acknowledges the financial support from the Council of Scientific and Industrial Research (CSIR), Government of India, New Delhi, through a NET Junior Research Fellowship vide grant No. 09/0059(15346)/2022-EMR-I. The work of BK has been supported in part by the Polish National Science Center (NCN) under grant 2020/37/B/ST2/02371, the Freedom of Research, Mobility of Science, and the Research Excellence Initiative of the University of Silesia in Katowice.


\appendix

\section{Product Rules of $A_4$ \label{C}}

The discrete group $A_4$ is a subgroup of $SU(3)$ which has 12 elements. The group has four irreducible representations, out of which three are one-dimensional and one is three-dimensional. In T-basis, the product of two triplets gives,

\begin{equation}
3\times3=1+1'+1''+3_S+3_A,
\end{equation}

where,

\begin{eqnarray}
1&=&(a_1 b_1+a_2 b_3+a_3 b_2)\\
1'&=&(a_3 b_3+a_1 b_2+a_2 b_1)\\
1''&=&(a_2 b_2+ a_1 b_3+ a_3 b_1)\\
(3\times3)_S&=& \frac{1}{3}\begin{bmatrix}
2 a_1 b_1-a_2 b_3-a_3 b_2\\
2 a_3 b_3-a_1 b_2-a_2 b_1\\
2 a_2 b_2-a_1 b_3-a_3 b_1\\
\end{bmatrix},\\
(3\times3)_A&=&\frac{1}{2}\begin{bmatrix}
a_2 b_3-a_3 b_2\\
a_1 b_2-a_2 b_1\\
a_3 b_1-a_1 b_3\\
\end{bmatrix}.
\end{eqnarray}

The trivial singlet can be obtained from the following singlet product rules,

\begin{equation}
1\times1=1,\,\,1'\times1''=1,\,\,1''\times1'=1.
\end{equation}

\section{The Scalar Potential \label{b}}

The $A_4 \times Z_{10} \times Z_2$ invariant scalar potential can be constructed in the following manner,
\begin{eqnarray}
V &=& V(\Phi)+V(H)+V(\Delta_1)+V(\Delta_2)+V(\eta)+V(\xi)+V(\chi)+V(\psi)+V(\kappa)\nonumber\\&&+V(\zeta)+V(\Phi H)+V(\Phi \Delta)+V(\Phi \Delta_2)+V(\Phi \eta)+V(\Phi \xi)+V(\Phi \chi)+V(\Phi \psi)\nonumber\\&&+V(\Phi \kappa)+V(\Phi \zeta)+V(H \Delta)+V(H \Delta_2)+V(H \eta)+V(H \xi)+V(H \chi)+\nonumber\\&&V(H \psi)+V(H \kappa)+V(H \zeta)+V(\Delta_1 \Delta_2)+V(\Delta_1 \eta)+V(\Delta_1\xi)+V(\Delta_1 \chi)+V(\Delta_1 \psi)\nonumber\\&&+V(\Delta_1 \kappa)+V(\Delta_1 \zeta)+V(\Delta_2 \eta)+V(\Delta_2 \xi)+V(\Delta_2 \chi)+V(\Delta_2 \psi)+V(\Delta_2 \kappa)+V(\Delta_2 \zeta)\nonumber\\&&+V(\eta \xi)+V(\eta \chi)+V(\eta \psi)+V(\eta \kappa)+V(\eta \zeta)+V(\xi \chi)+V(\xi \psi)+V(\xi \kappa)+\nonumber\\&&V(\xi \zeta)+V(\chi \psi)+V(\chi \kappa)+V(\chi \zeta)++V(\psi \kappa)+V(\psi \zeta)+V(\kappa \zeta),
\end{eqnarray}
where,

\begin{eqnarray}
V(\Phi)&=& \mu^2_\Phi (\Phi^\dagger \Phi)+\lambda^\Phi_1 (\Phi^\dagger \Phi) (\Phi^\dagger \Phi)+\lambda^\Phi_2 (\Phi^\dagger \Phi)_{1'} (\Phi^\dagger \Phi)_{1''}+\lambda^\Phi_3 (\Phi^\dagger \Phi)_{3_s} (\Phi^\dagger \Phi)_{3_s}+\lambda^\Phi_4 (\Phi^\dagger \Phi)_{3_s}\nonumber\\&& (\Phi^\dagger \Phi)_{3_a}+\lambda^\Phi_5 (\Phi^\dagger \Phi)_{3_a} (\Phi^\dagger \Phi)_{3_a},\\
V(H)&=& \mu^2_H (H^\dagger H)+\lambda^H (H^\dagger H)(H^\dagger H),\\
V(\Delta_1)&=& \mu^2_{\Delta_1} Tr(\Delta_1^\dagger \Delta_1)+\lambda^\Delta_1 \,Tr(\Delta_1^\dagger \Delta_1)Tr(\Delta_1^\dagger \Delta_1),\\
V(\Delta_2)&=& \mu^2_{\Delta_2} Tr(\Delta_2^\dagger \Delta_2)+\lambda^{\Delta_2}\, Tr(\Delta_2^\dagger \Delta_2)Tr(\Delta_2^\dagger \Delta_2),\\
V(\eta)&=& \mu^2_\eta (\eta^\dagger \eta)+\lambda^\eta (\eta^\dagger \eta)(\eta^\dagger \eta),\\
V(\xi)&=& \mu^2_\xi (\xi^\dagger \xi)+\lambda^\xi (\xi^\dagger \xi)(\xi^\dagger \xi),\\
V(\chi)&=& \mu^2_\chi (\chi^\dagger \chi)+\lambda^\chi (\chi^\dagger \chi)(\chi^\dagger \chi),\\
V(\zeta)&=& \mu^2_\zeta (\zeta^\dagger \zeta)+\lambda^\zeta (\zeta^\dagger \zeta)(\zeta^\dagger \zeta),\\
V(\psi)&=& \mu^2_\psi (\psi^\dagger \psi)+\lambda^\psi_1 (\psi^\dagger \psi) (\psi^\dagger \psi)+\lambda^\psi_2 (\psi^\dagger \psi)_{1'} (\psi^\dagger \psi)_{1''}+\lambda^\psi_3 (\psi^\dagger \psi)_{3_s} (\psi^\dagger \psi)_{3_s}+\lambda^\psi_4 (\psi^\dagger \psi)_{3_s}\nonumber\\&& (\psi^\dagger \psi)_{3_a}+\lambda^\psi_5 (\psi^\dagger \psi)_{3_a} (\psi^\dagger \psi)_{3_a},
\end{eqnarray}

\begin{eqnarray}
V(\kappa)&=& \mu^2_\kappa (\kappa^\dagger \kappa)+\lambda^\kappa_1 (\kappa^\dagger \kappa) (\kappa^\dagger \kappa)+\lambda^\kappa_2 (\kappa^\dagger \kappa)_{1'} (\kappa^\dagger \kappa)_{1''}+\lambda^\kappa_3 (\kappa^\dagger \kappa)_{3_s} (\kappa^\dagger \kappa)_{3_s}+\lambda^\kappa_4 (\kappa^\dagger \kappa)_{3_s}\nonumber\\&& (\kappa^\dagger \kappa)_{3_a}+\lambda^\kappa_5 (\kappa^\dagger \kappa)_{3_a} (\kappa^\dagger \kappa)_{3_a},\\
V(\Phi H)&=& \lambda^{\Phi H}_1 (\Phi^\dagger \Phi)(H^\dagger H),\\
V(\Phi \Delta_1)&=& \lambda^{\Phi \Delta_1}_1 (\Phi^\dagger \Phi)\,Tr(\Delta_1^\dagger \Delta_1),\\
V(\Phi \Delta_2)&=& \lambda^{\Phi \Delta_2}_1 (\Phi^\dagger \Phi)\,Tr(\Delta_2^\dagger \Delta_2),\\
V(\Phi \eta)&=& \lambda^{\Phi \eta}_1 (\Phi^\dagger \Phi)(\eta^\dagger \eta),\\
V(\Phi \xi)&=& \lambda^{\Phi \xi}_1 (\Phi^\dagger \Phi)(\xi^\dagger \xi),\\
V(\Phi \chi)&=& \lambda^{\Phi \chi}_1 (\Phi^\dagger \Phi)(\chi^\dagger \chi),\\
V(\Phi \zeta)&=& \lambda^{\Phi \zeta}_1 (\Phi^\dagger \Phi)(\zeta^\dagger \zeta),\\
V(\Phi \psi)&=& \lambda^{\Phi \psi}_1 (\Phi^\dagger \Phi)(\psi^\dagger \psi)+\lambda^{\Phi \psi}_2 [(\Phi^\dagger \Phi)_{1'}(\psi^\dagger \psi)_{1''}+(\Phi^\dagger \Phi)_{1''}(\psi^\dagger \psi)_{1'}]+\lambda^{\Phi \psi}_3 \nonumber\\&& (\Phi^\dagger \Phi)_{3_s}(\psi^\dagger \psi)_{3_s}+\lambda^{\Phi \psi}_4 [(\Phi^\dagger \Phi)_{3_s}(\psi^\dagger \psi)_{3_a}+(\Phi^\dagger \Phi)_{3_a}(\psi^\dagger \psi)_{3_s}]+\lambda^{\Phi \psi}_5 \nonumber\\&& (\Phi^\dagger \Phi)_{3_a}(\psi^\dagger \psi)_{3_a},\\
V(\Phi \kappa)&=& \lambda^{\Phi \kappa}_1 (\Phi^\dagger \Phi)(\kappa^\dagger \kappa)+\lambda^{\Phi \kappa}_2 [(\Phi^\dagger \Phi)_{1'}(\kappa^\dagger \kappa)_{1''}+(\Phi^\dagger \Phi)_{1''}(\kappa^\dagger \kappa)_{1'}]+\lambda^{\Phi \kappa}_3  (\Phi^\dagger \Phi)_{3_s}\nonumber\\&&(\kappa^\dagger \kappa)_{3_s}+\lambda^{\Phi \kappa}_4 [(\Phi^\dagger \Phi)_{3_s}(\kappa^\dagger \kappa)_{3_a}+(\Phi^\dagger \Phi)_{3_a}(\kappa^\dagger \kappa)_{3_s}]+\lambda^{\Phi \kappa}_5 (\Phi^\dagger \Phi)_{3_a}(\kappa^\dagger \kappa)_{3_a},\\
V(H \Delta_1)&=& \lambda^{H \Delta_1}_1 (H^\dagger H)\,Tr(\Delta_1^\dagger \Delta_1)+\frac{1}{\Lambda}(H^T i \sigma_2 \Delta_1^\dagger H \Phi \psi),\\
V(H \Delta_2)&=& \lambda^{H \Delta_2}_1 (H^\dagger H)\,Tr(\Delta_2^\dagger \Delta_2)+\frac{1}{\Lambda}(H^T i \sigma_2 \Delta_2^\dagger H \Phi \psi),\\
V(H \eta)&=& \lambda^{H \eta}_1 (H^\dagger H)(\eta^\dagger \eta),\\
V(H \xi)&=& \lambda^{H \xi}_1 (H^\dagger H)(\xi^\dagger \xi),\\
V(H \chi)&=& \lambda^{H \chi}_1 (H^\dagger H)(\chi^\dagger \chi),\\
V(H \psi)&=& \lambda^{H \psi}_1 (H^\dagger H)(\psi^\dagger \psi),\\
V(H \kappa)&=& \lambda^{H \kappa}_1 (H^\dagger H)(\kappa^\dagger \kappa),\\
V(H \zeta)&=& \lambda^{H \zeta}_1 (H^\dagger H)(\zeta^\dagger \zeta),\\
V(\Delta_1 \Delta_2)&=& \lambda^{\Delta_1 \Delta_2}_1 (\Delta_1^\dagger \Delta_1)\,Tr(\Delta_2^\dagger \Delta_2),\\
V(\Delta_1 \eta)&=& \lambda^{\Delta_1 \eta}_1\,Tr (\Delta_1^\dagger \Delta_1)(\eta^\dagger \eta),\\
V(\Delta_1 \xi)&=& \lambda^{\Delta_1 \xi}_1 \,Tr(\Delta_1^\dagger \Delta_1)(\xi^\dagger \xi),\\
V(\Delta_1 \chi)&=& \lambda^{\Delta_1 \chi}_1 \,Tr(\Delta_1^\dagger \Delta_1)(\chi^\dagger \chi),\\
V(\Delta_1 \psi)&=& \lambda^{\Delta_1 \psi}_1 \,Tr(\Delta_1^\dagger \Delta_1)(\psi^\dagger \psi),\\
V(\Delta_1 \kappa)&=& \lambda^{\Delta_1 \kappa}_1 \,Tr(\Delta_1^\dagger \Delta_1)(\kappa^\dagger \kappa),\\
V(\Delta_1 \zeta)&=& \lambda^{\Delta_1 \zeta}_1 \,Tr(\Delta_1^\dagger \Delta_1)(\zeta^\dagger \zeta),\\
V(\Delta_2 \eta)&=& \lambda^{\Delta_2 \eta}_1\,Tr (\Delta_2^\dagger \Delta_2)(\eta^\dagger \eta),\\
V(\Delta_2 \xi)&=& \lambda^{\Delta_2 \xi}_1 \,Tr(\Delta_2^\dagger \Delta_2)(\xi^\dagger \xi),\\
V(\Delta_2 \chi)&=& \lambda^{\Delta_2 \chi}_1 \,Tr(\Delta_2^\dagger \Delta_2)(\chi^\dagger \chi),\\
V(\Delta_2 \psi)&=& \lambda^{\Delta_2 \psi}_1 \,Tr(\Delta_2^\dagger \Delta_2)(\psi^\dagger \psi),\\
V(\Delta_2 \kappa)&=& \lambda^{\Delta_2 \kappa}_1 \,Tr(\Delta_2^\dagger \Delta_2)(\kappa^\dagger \kappa),\\
V(\Delta_2 \zeta)&=& \lambda^{\Delta_2 \zeta}_1 \,Tr(\Delta_2^\dagger \Delta_2)(\zeta^\dagger \zeta),\\
V(\eta \xi)&=& \lambda^{\eta \xi}_1 (\eta^\dagger \eta)(\xi^\dagger \xi),\\
V(\eta \chi)&=& \lambda^{\eta \chi}_1 (\eta^\dagger \eta)(\chi^\dagger \chi),\\
V(\eta \psi)&=& \lambda^{\eta \psi}_1 (\eta^\dagger \eta)(\psi^\dagger \psi),\\
V(\eta \kappa)&=& \lambda^{\eta \kappa}_1 (\eta^\dagger \eta)(\kappa^\dagger \kappa),\\
V(\eta \zeta)&=& \lambda^{\eta \zeta}_1 (\eta^\dagger \eta)(\zeta^\dagger \zeta),\\
V(\xi \chi)&=& \lambda^{\xi \chi}_1 (\xi^\dagger \xi)(\chi^\dagger \chi),\\
V(\xi \psi)&=& \lambda^{\xi \psi}_1 (\xi^\dagger \xi)(\psi^\dagger \psi),
\end{eqnarray}

\begin{eqnarray}
V(\xi \kappa)&=& \lambda^{\xi \kappa}_1 (\xi^\dagger \xi)(\kappa^\dagger \kappa),\\
V(\xi \zeta)&=& \lambda^{\xi \zeta}_1 (\xi^\dagger \xi)(\zeta^\dagger \zeta),\\
V(\chi \psi)&=& \lambda^{\chi \psi}_1 (\chi^\dagger \chi)(\psi^\dagger \psi),\\
V(\chi \kappa)&=& \lambda^{\chi \kappa}_1 (\chi^\dagger \chi)(\kappa^\dagger \kappa),\\
V(\chi \zeta)&=& \lambda^{\chi \zeta}_1 (\chi^\dagger \chi)(\zeta^\dagger \zeta),\\
V( \psi \kappa)&=& \lambda^{ \psi \kappa}_1 (\psi^\dagger \psi)(\kappa^\dagger \kappa)+\lambda^{ \psi \kappa}_2 [(\psi^\dagger \psi)_{1'}(\kappa^\dagger \kappa)_{1''}+(\psi^\dagger \psi)_{1''}(\kappa^\dagger \kappa)_{1'}]+\lambda^{ \psi \kappa}_3 \nonumber\\&& (\psi^\dagger \psi)_{3_s}(\kappa^\dagger \kappa)_{3_s}+\lambda^{ \psi \kappa}_4 [(\psi^\dagger \psi)_{3_s}(\kappa^\dagger \kappa)_{3_a}+(\psi^\dagger \psi)_{3_a}(\kappa^\dagger \kappa)_{3_s}]+\lambda^{\psi \kappa}_5 \nonumber\\&& (\psi^\dagger \psi)_{3_a}(\kappa^\dagger \kappa)_{3_a},\\
V(\psi \zeta)&=& \lambda^{\psi \zeta}_1 (\psi^\dagger \psi)(\zeta^\dagger \zeta),\\
V(\kappa \zeta)&=& \lambda^{\kappa \zeta}_1 (\kappa^\dagger \kappa)(\zeta^\dagger \zeta),
\end{eqnarray}

 The following equations hold good for the chosen vacuum expectation values (VEVs) $\langle \Phi \rangle_{\circ}=v_{\Phi}\,(1,0,0)^{T}$, $\langle H \rangle_{\circ}=v_{H}$, $\langle\eta\rangle_{\circ}=v_{\eta}$, $\langle\chi\rangle_{\circ}=v_{\chi}$, $\langle\xi\rangle_{\circ}=v_{\xi}$, $\langle\varsigma\rangle_{\circ}=v_{\varsigma}$, $\langle\Delta_1\rangle_{\circ}=v_{\Delta_1}$, $\langle\Delta_2\rangle_{\circ}=v_{\Delta_2}$, $\langle\psi\rangle_{\circ}=v_{\psi}(0,1,-1)^{T}$ and $\langle\kappa\rangle_{\circ}=v_{\kappa}(0,1,-1)^{T}$, based on the minimization conditions of the scalar potential:
 
 \begin{eqnarray}
 \frac{\partial V}{\partial \Phi_1}&=& \frac{1}{9} v_\Phi (-\frac{24}{\Lambda} \mu_2 v_{\Delta_1} v_H^2 v_\chi^2 \lambda _1^{\Phi \chi }+9 v_{\Delta_2}^2 \lambda _1^{\Phi \Delta_2 }+9 v_\eta^2 \lambda _1^{\Phi \eta }+9 v_H^2 \lambda _1^{\Phi H}+9 v_\kappa^2 \lambda _1^{\Phi \kappa }+4 v_\kappa^2 \lambda _3^{\Phi \kappa }\nonumber\\&&+9 \mu _{\Phi }^2-18 v_\psi^2 \lambda _1^{\Phi \psi} +4 v_\psi^2 \lambda _3^{\Phi \psi }+2 v_\Phi^2 \left(9 \lambda _1^{\Phi }+4 \lambda _3^{\Phi }\right)+9 \lambda _1^{\Phi \xi } v_\xi^2+9 \lambda _1^{\phi \zeta } v_\zeta^2\nonumber\\&&+9 v_{\Delta_1}^2 \lambda_1^{\Phi \Delta_1})=0,\\
 \frac{\partial V}{\partial \Phi_2}&=& \frac{1}{9} v_\Phi ( \frac{12}{\Lambda} v_H^2 v_{\Delta_1 }(\mu _2 v_{\kappa }-\mu _1 v_{\psi })+18 v_\kappa^2 \lambda _2^{\Phi \kappa }+2 v_\kappa^2 \lambda _3^{\Phi \kappa }-3 v_\kappa^2 \lambda _4^{\Phi \kappa }+9 v_\psi^2 \lambda _2^{\Phi \psi }\nonumber\\&&-2 v_\psi^2 \lambda _3^{\Phi \psi }+3 v_\psi^2 \lambda _4^{\Phi \psi })=0,\\
 \frac{\partial V}{\partial \Phi_3}&=& \frac{1}{9} v_\Phi (v_\kappa^2 (9 \lambda _2^{\Phi \kappa }-2 \lambda _3^{\Phi \kappa }-3 \lambda _4^{\Phi \kappa })+v_\psi^2 (\frac{12}{\Lambda^2}\mu _1 v_H^2 v_{\Delta_1 }+ 9 \lambda _2^{\Phi \psi }-2 \lambda _3^{\Phi \psi }\nonumber\\&&-3 \lambda _4^{\Phi \psi }))=0,\\
 \frac{\partial V}{\partial \psi_1}&=& -\frac{4}{3 \Lambda} \mu _1 v_H^2 v_{\Delta_1 } v_{\Phi }^2+ \frac{1}{9} v_\psi (v_\kappa^2 (9 \lambda _2^{\kappa \psi }+4 \lambda _3^{\kappa \psi })-6 v_\psi^2 \lambda _4^{\psi })=0,\\
 \frac{\partial V}{\partial \psi_2}&=& \frac{1}{9} v_\psi (-9 v_H^2 \lambda _1^{H\psi}-9 \mu _{\psi }^2-9 \lambda _1^{\Delta_1 \psi } v_{\Delta_1 }^2-9 v_{\Delta_2 }^2 \lambda _1^{\Delta_2 \psi }-9 \lambda _1^{\psi \zeta } v_{\zeta }^2-9 \lambda _1^{\eta \psi } v_{\eta }^2-9 \lambda _1^{\kappa \psi } v_{\kappa }^2\nonumber\\&&+9 \lambda _2^{\kappa \psi } v_{\kappa }^2+6 \lambda _3^{\kappa \psi } v_{\kappa }^2+3 \lambda _4^{\kappa \psi } v_{\kappa }^2-9 \lambda _1^{\xi \psi } v_{\xi }^2-9 \lambda _1^{\Phi \psi } v_{\Phi }^2+2 \lambda _3^{\Phi \psi } v_{\Phi }^2+3 \lambda _4^{\Phi \psi } v_{\Phi }^2\nonumber\\&&-9 \lambda _1^{\chi \psi } v_{\chi }^2+36 \lambda _1^{\psi } v_{\psi }^2+9 \lambda _2^{\psi } v_{\psi }^2+12 \lambda _3^{\psi } v_{\psi }^2+3 \lambda _4^{\psi } v_{\psi }^2)=0,\\
 \frac{\partial V}{\partial \psi_3}&=& \frac{1}{9} v_\psi (9 v_H^2 \lambda _1^{H\psi }+9 \mu _{\psi }^2+9 \lambda _1^{\Delta_1 \psi } v_{\Delta_1 }^2+9 v_{\Delta_2}^2 \lambda _1^{\Delta_2 \psi }+9 \lambda _1^{\psi \zeta } v_{\zeta }^2+9 \lambda _1^{\eta \psi } v_{\eta }^2+9 \lambda _1^{\kappa \psi } v_{\kappa }^2\nonumber\\&&-18 \lambda _2^{\kappa \psi } v_{\kappa }^2+2 \lambda _3^{\kappa \psi } v_{\kappa }^2+3 \lambda _4^{\kappa \psi } v_{\kappa }^2+9 \lambda _1^{\xi \psi } v_{\xi }^2-2 \lambda _3^{\Phi \psi } v_{\Phi }^2+3 \lambda _4^{\Phi \psi } v_{\Phi }^2+9 \lambda _1^{\Phi \psi } v_{\psi }^2\nonumber\\&&+9 \lambda _1^{\chi \psi } v_{\chi }^2-36 \lambda _1^{\psi } v_{\psi }^2-9 \lambda _2^{\psi } v_{\psi }^2-12 \lambda _3^{\psi } v_{\psi }^2+3 \lambda _4^{\psi } v_{\psi }^2)=0,\\
  \frac{\partial V}{\partial \kappa_1}&=& -\frac{4}{3 \Lambda}\mu _2 v_H^2 v_{\Delta_1 } v_{\Phi }^2+ \lambda _1^{\Delta_1 \kappa} v_{\Delta_1 }^2 v_{\kappa }+\frac{1}{9} v_\kappa (9 v_H^2 \lambda _1^{H \kappa }+9 \mu _{\kappa }^2+9 v_{\Delta_2}^2 \lambda _1^{\Delta_2 \kappa }+9 \lambda _1^{\eta \kappa } v_{\eta }^2+9 \lambda _2^{\kappa \psi } v_{\kappa }^2\nonumber\\&&+18 \lambda _1^{\kappa } v_{\kappa }^2+9 \lambda _2^{\kappa } v_{\kappa }^2+4 \lambda _3^{\kappa } v_{\kappa }^2+3 \lambda _4^{\kappa } v_{\kappa }^2+9 \lambda _1^{\kappa \zeta } v_{\xi }^2-18 \lambda _1^{\kappa \psi } v_{\psi }^2+2 \lambda _3^{\kappa \psi } v_{\psi }^2+3 \lambda _4^{\kappa \psi } v_{\psi }^2\nonumber\\&&+9 \lambda _1^{\xi \kappa } v_{\xi }^2+9 \lambda _1^{\Phi \kappa } v_{\Phi }^2+4 \lambda _3^{\Phi \kappa } v_{\Phi }^2+9 \lambda _1^{\chi \kappa } v_{\chi }^2)=0,\\
  \frac{\partial V}{\partial \kappa_2}&=&\frac{1}{9} v_\kappa(9 v_H^2 \lambda _1^{H \kappa }+9 \mu _{\kappa }^2+9 \lambda _1^{\Delta_1  \kappa} v_{\Delta_1 }^2+9 v_{\Delta_2}^2 \lambda _1^{\Delta_2 \kappa }+9 \lambda _1^{\kappa \zeta } v_{\zeta }^2+9 \lambda _1^{\eta \kappa } v_{\eta }^2+18 \lambda _1^{\kappa } v_{\kappa }^2\nonumber\\&&+18 \lambda _2^{\kappa } v_{\kappa }^2-6 \lambda _4^{\kappa } v_{\kappa }^2-18 \lambda _1^{\kappa \psi } v_{\psi }^2+9 \lambda _2^{\kappa \psi } v_{\psi }^2-4 \lambda _3^{\kappa \psi } v_{\psi }^2+9 \lambda _1^{\xi \kappa } v_{\xi }^2+9 \lambda _1^{\Phi \kappa } v_{\Phi }^2\nonumber\\&&-2 \lambda _3^{\Phi \kappa } v_{\Phi }^2-3 \lambda _4^{\phi \kappa } v_{\Phi }^2+9 \lambda _1^{\chi \kappa } v_{\chi }^2)=0,\\
  \frac{\partial V}{\partial \kappa_3}&=&\frac{1}{9} v_\kappa(27 \lambda _2^{\kappa } v_{\kappa }^2-12 \lambda _3^{\kappa } v_{\kappa }^2+18 \lambda _2^{\kappa \psi } v_{\psi }^2+2 \lambda _3^{\kappa \psi } v_{\psi }^2-3(\lambda _4^{\kappa } v_{\kappa }^2+\lambda _4^{\kappa \psi } v_{\psi }^2))=0,
 \end{eqnarray}
 
\begin{eqnarray}
\frac{\partial V}{\partial H}&=&\frac{1}{9} v_H (-\frac{8}{\Lambda} \mu _2 v_{\Delta_1 } v_{\kappa } v_{\Phi }^2+3 \lambda _1^{H \Delta_1 } v_{\Delta_1 }^2+3(\mu _H^2+v_{\Delta_2}^2 \lambda _1^{\Delta_2 H}+v_{\kappa }^2 \lambda _1^{H \kappa }+2 \lambda ^H v_H^2+\lambda _1^{H\zeta } v_{\zeta }^2\nonumber\\&&+\lambda _1^{H\eta } v_{\eta }^2+\lambda _1^{H\xi } v_{\xi }^2+\lambda _1^{H\chi } v_{\chi }^2-2 \lambda _1^{H\psi } v_{\psi }^2+\lambda _1^{\Phi H} v_{\Phi }^2))=0,\\
  \frac{\partial V}{\partial \Delta_1}&=&2 \lambda ^{\Delta_1 } v_{\Delta_1 }^3-\frac{4}{\Lambda} \mu _2 v_{H} v_{\kappa } v_{\Phi }^2+v_{\Delta_1}(\mu _{\Delta_1 }^2+v_H^2 \lambda _1^{\Delta_1  H}+v_{\kappa }^2 \lambda _1^{\Delta_1  \kappa }+\lambda_1^{\Delta_1  \Delta_2} v_{\Delta_2}^2+\lambda _1^{\Delta_1 \zeta } v_{\zeta }^2+\lambda _1^{\Delta_1 \eta } v_{\eta }^2\nonumber\\&&+\lambda _1^{\Delta_1 \xi } v_{\xi }^2+\lambda _1^{\Delta_1 \chi } v_{\chi }^2-2 \lambda _1^{\Delta_1 \psi } v_{\psi }^2+\lambda _1^{\Phi \Delta_1 } v_{\Phi }^2)=0,\\
  \frac{\partial V}{\partial \Delta_2}&=&v_{\Delta_2}(\mu _{\Delta_2}^2+v_H^2 \lambda _1^{\Delta_2 H}+v^2{}_{\Phi } \lambda _1^{\Delta_2 \phi }+v_{\zeta }^2 \lambda _1^{\Delta_2 \zeta }+v_{\eta }^2 \lambda _1^{\Delta_2 \eta }+v_{\kappa }^2 \lambda _1^{\Delta_2 \kappa }+2 \lambda ^{\Delta_2} v_{\Delta_2}^2\nonumber\\&&+v_{\xi }^2 \lambda _1^{\Delta_2 \xi }+v_{\chi }^2 \lambda _1^{\Delta_2 \chi }-2 v_{\psi }^2 \lambda _1^{\Delta_2 \psi }+\lambda _1^{\Delta_1 \Delta_2} v_{\Delta_1 }^2)=0,\\
\frac{\partial V}{\partial \eta}&=&v_\eta (\mu _{\eta }^2+v_H^2 \lambda _1^{\text{H$\eta $}}+\lambda _1^{\Delta_1 \eta } v_{\Delta_1 }^2+v_{\Delta_2}^2 \lambda _1^{\Delta_2 \eta }+\lambda _1^{\eta \zeta } v_{\zeta }^2+2 \lambda ^{\eta } v_{\eta }^2+\lambda _1^{\eta \kappa } v_{\kappa }^2+\lambda _1^{\eta \xi } v_{\xi }^2\nonumber\\&&+\lambda _1^{\eta \chi } v_{\chi }^2-2 \lambda _1^{\eta \psi } v_{\psi }^2+\lambda _1^{\Phi \eta } v_{\Phi }^2)=0,\\
\frac{\partial V}{\partial \xi}&=&v_\xi (v_H^2 \lambda _1^{H\xi}+\mu _{\xi }^2+\lambda _1^{\Delta_1 \xi } v_{\Delta_1 }^2+v_{\Delta_2}^2 \lambda _1^{\Delta_2 \xi }+\lambda _1^{\xi \zeta } v_{\zeta }^2+\lambda _1^{\eta \xi } v_{\eta }^2+\lambda _1^{\xi \kappa } v_{\kappa }^2+2 \lambda ^{\xi } v_{\xi }^2\nonumber\\&&+\lambda _1^{\xi \chi } v_{\chi }^2-2 \lambda _1^{\xi \psi } v_{\psi }^2+\lambda _1^{\Phi \xi } v_{\Phi }^2)=0,\\
\frac{\partial V}{\partial \chi}&=&v_\chi(v_H^2 \lambda _1^{H\chi }+\mu _{\chi }^2+\lambda _1^{\Delta_1 \chi } v_{\Delta_1 }^2+v_{\Delta_2}^2 \lambda _1^{\Delta_2 \chi }+\lambda _1^{\chi \zeta } v_{\zeta }^2+\lambda _1^{\eta \chi } v_{\eta }^2+\lambda _1^{\chi \kappa } v_{\kappa }^2+2 \lambda ^{\chi } v_{\chi }^2\nonumber\\&&+\lambda _1^{\xi \chi } v_{\xi }^2+\lambda _1^{\Phi \chi } v_{\Phi }^2-2 \lambda _1^{\chi \psi } v_{\psi }^2)=0,\\
\frac{\partial V}{\partial \zeta}&=&v_\zeta (\mu _{\zeta }^2+v_H^2 \lambda _1^{H\zeta }+\lambda _1^{\Delta_1 \zeta } v_{\Delta_1 }^2+v_{\Delta_2}^2 \lambda _1^{\Delta_2 \zeta }+2 \lambda ^{\zeta } v_{\zeta }^2+\lambda _1^{\eta \zeta } v_{\eta }^2+\lambda _1^{\kappa \zeta } v_{\kappa }^2+\lambda _1^{\xi \zeta } v_{\xi }^2\nonumber\\&&+\lambda _1^{\Phi \zeta } v_{\Phi }^2+\lambda _1^{\chi \zeta } v_{\chi }^2-2 \lambda _1^{\psi \zeta } v_{\psi }^2)=0.
\end{eqnarray}

\biboptions{sort&compress}
 \bibliography{ref.bib}
\end{document}